\documentclass[12pt]{article}
\pdfoutput=1
\usepackage[nosort]{cite}

\usepackage{xcolor}

\usepackage{epsfig}
\usepackage{amsfonts}
\usepackage{amscd}
\usepackage{latexsym}
\usepackage{amsmath,amssymb}
\usepackage{verbatim}
\usepackage{setspace}
\usepackage{color}
\usepackage{fancyhdr}
\usepackage{cite}
\usepackage{hyperref}
\usepackage{tikz}
\usepackage{slashed}
\usepackage{soul}
\usepackage{multirow}
\usepackage{makecell}
\usetikzlibrary{calc}
\usetikzlibrary{topaths}
\usetikzlibrary{decorations}
\usetikzlibrary{decorations.pathmorphing}

\usepackage{draft}
\usepackage{hyperref}
\usepackage{graphicx,color,subfig}
\usepackage{cite}
\usepackage{mciteplus}
\usepackage{skak}

\numberwithin{equation}{section}

\newcommand{\lcm}{\operatorname{lcm}}
\renewcommand{\b}[1]{\boldsymbol{#1}}
\newcommand{\diag}{\operatorname{diag}}

\newcommand{\Jac}{\operatorname{Jac}}

\def\({\left(}
\def\){\right)}

\begin{document}

\begin{titlepage}
	
\begin{center}

\hfill \\
\hfill YITP-SB-2022-33, MIT/CTP-5471\\
\vskip 1cm

\title{Gapped Lineon and Fracton Models on Graphs}

\author{Pranay Gorantla$^{1}$, Ho Tat Lam$^{2}$, Nathan Seiberg$^{3}$ and Shu-Heng Shao$^{4}$}

\address{${}^{1}$Physics Department, Princeton University}
\address{${}^{2}$Center for Theoretical Physics, Massachusetts Institute of Technology}
\address{${}^{3}$School of Natural Sciences, Institute for Advanced Study}
\address{${}^{4}$C.\ N.\ Yang Institute for Theoretical Physics, Stony Brook University}

\end{center}

\vspace{2.0cm}

\begin{abstract}\noindent
We introduce a  $\mathbb{Z}_N$ stabilizer code that can be defined on any spatial lattice of the form $\Gamma\times C_{L_z}$, where $\Gamma$ is a general graph.
We also present  the low-energy limit of this stabilizer code as  a Euclidean lattice action, which we refer to as  the  anisotropic  $\mathbb{Z}_N$ Laplacian model.
It is gapped, robust (i.e., stable under small deformations), and has lineons.
Its ground state degeneracy (GSD) is expressed in terms of a ``mod $N$-reduction'' of the Jacobian group of the graph $\Gamma$.
In the special case when space is an $L\times L\times L_z$ cubic lattice, the logarithm of the GSD depends on $L$ in an erratic way and grows no faster than $O(L)$. We also discuss another gapped model, the $\mathbb{Z}_N$ Laplacian model, which can be defined on any graph. It has  fractons and a similarly strange GSD.

\end{abstract}

\vfill
	
\end{titlepage}

\eject

\tableofcontents

\section{Introduction}

In recent years, there has been a rapid development in the study of exotic lattice models in condensed matter systems. Some models, known as fractons \cite{Chamon:2004lew,Haah:2011drr,Vijay:2016phm} exhibit a variety of surprising features.
These include a robust ground state degeneracy (GSD) that grows sub-extensively in the system size \cite{Haah:2020ghp}, as well as particle excitations with restricted mobility.
Many of these unusual properties can be understood as following from their exotic global symmetries.  These symmetries are also   the underlying reasons why these lattice models defy a conventional continuum limit.
See \cite{Nandkishore:2018sel,Pretko:2020cko,Grosvenor:2021hkn,Brauner:2022rvf,McGreevy:2022oyu,Cordova:2022ruw} for reviews on these novel topological phases of matter and their exotic global symmetries.

Most of these exotic lattice models are defined on a cubic lattice, or lattices with additional structure such as foliation \cite{Slagle:2018wyl,Shirley:2017suz,Shirley:2018nhn,Shirley:2018hkm,Shirley:2018vtc,Slagle:2018kqf,Slagle:2018swq,
Shirley:2019uou,Slagle:2020ugk,Hsin:2021mjn,Geng:2021cmq}.
It is then natural to ask if there are exotic models that can be defined on a general lattice graph.
Recently, two such lattice models, the Laplacian $\phi$-theory and the $U(1)$ Laplacian gauge theory, were proposed in \cite{Gorantla:2022mrp,Gorantla:2022ssr} using the discrete Laplacian operator $\Delta_L$. (See also \cite{Manoj:2021rpq} for a model along this line.)
The former has a large GSD being the number of spanning trees of the spatial graph, which is a common measure of complexity, but it does not have fractons.
The latter has defects representing immobile fracton particles, but it has  no large GSD.

\begin{figure}[t]
\begin{center}
\includegraphics[scale=0.3]{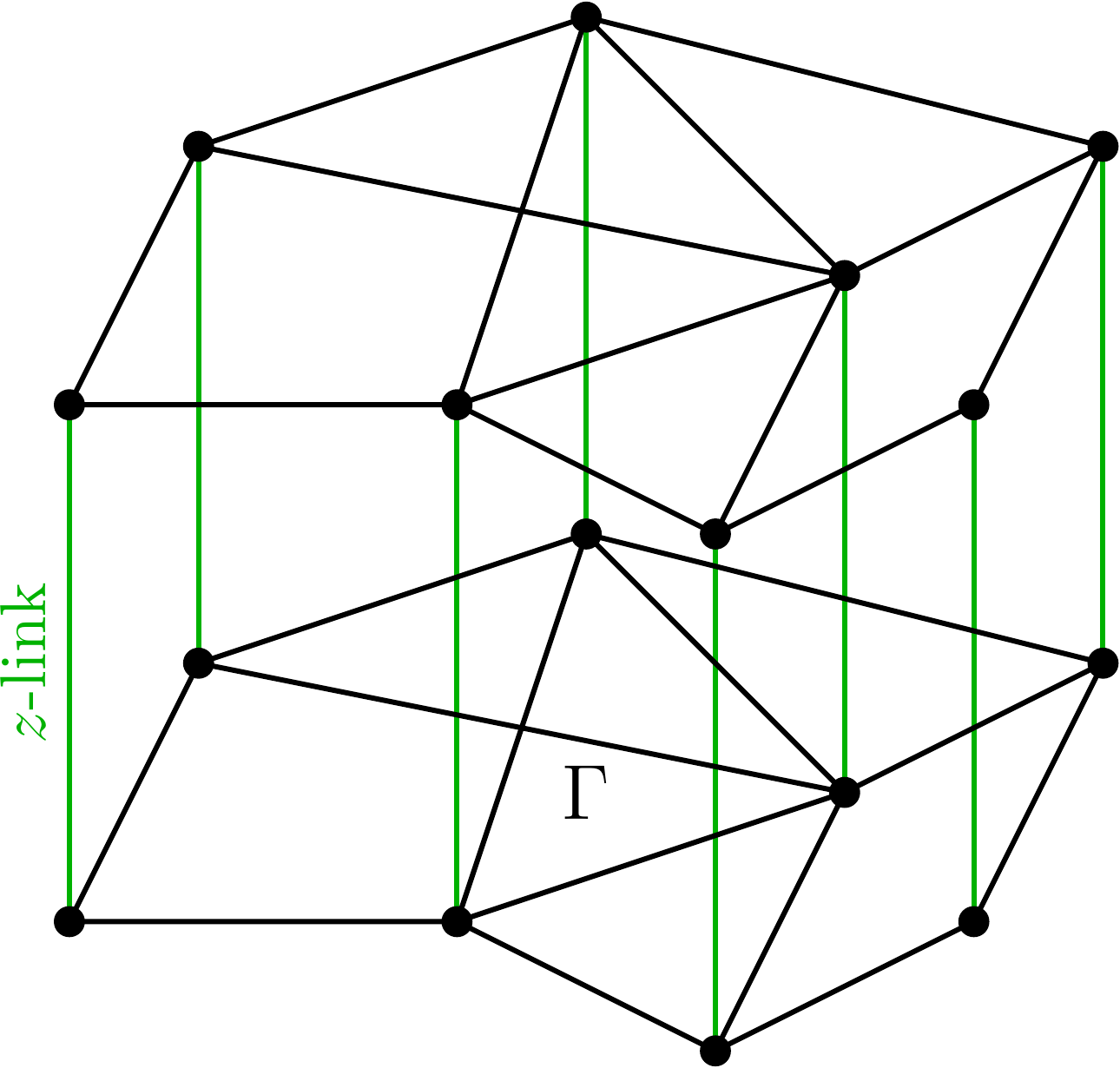}
\end{center}
\caption{The spatial lattice $\Gamma \times C_{L_z}$: the black lines correspond to the edges of the graph $\Gamma$, and the green lines represent the $z$-links between two copies of $\Gamma$. Each site of the lattice is labelled as $(i,z)$, where $i$ denotes a vertex of the graph $\Gamma$ and $z$ denotes a vertex of $C_{L_z}$.}\label{fig:ZNanisolap-lat}
\end{figure}

\begin{figure}[t]
\begin{center}
\includegraphics[scale=0.6]{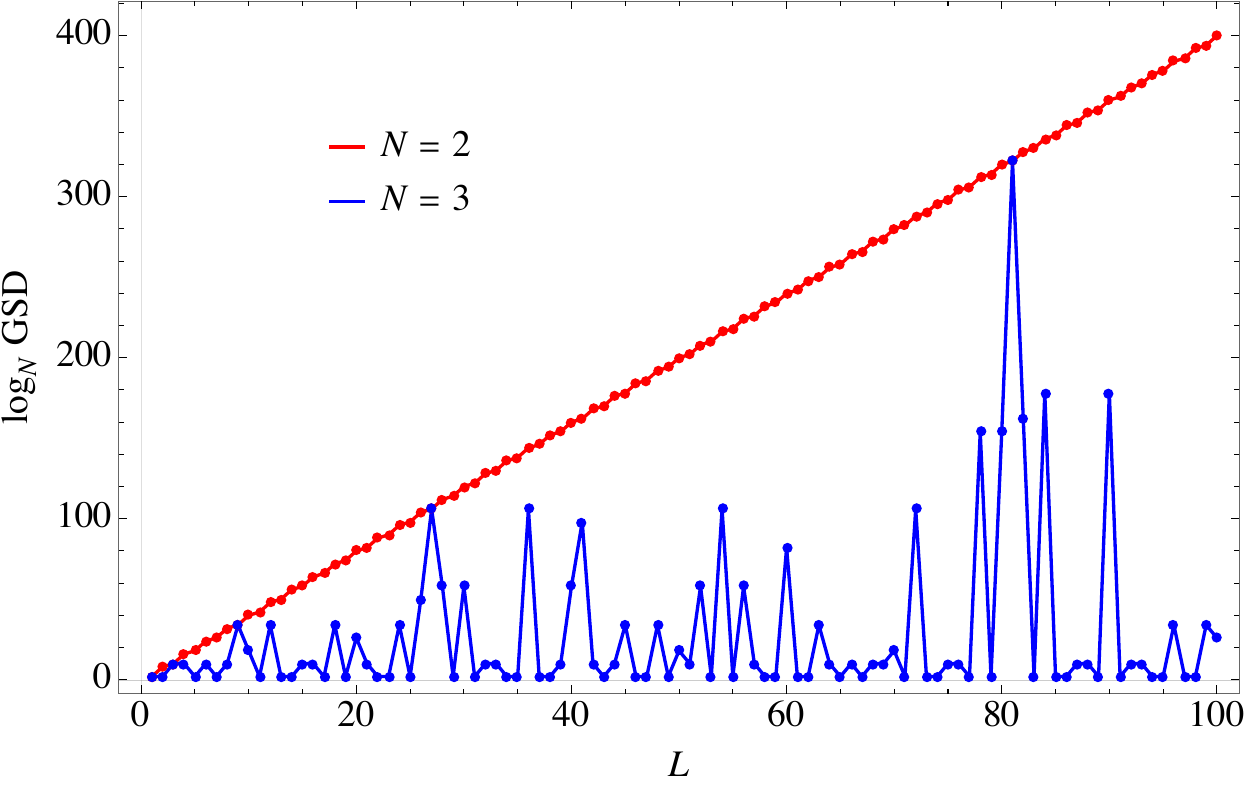}
\end{center}
\caption{The logarithm of the ground state degeneracy $\log_N\text{GSD}$ of the 3+1d anisotropic $\mathbb Z_N$ Laplacian model of lineons on a cubic lattice with $L\times L\times L_z$ sites (i.e., $\Gamma = C_L \times C_L$) for $N=2$ (red) and $N=3$ (blue), and $1\le L \le 100$. While $\log_N \text{GSD}$ grows steadily as $\sim 4L$ for $N=2$, it behaves erratically for $N=3$. In fact, for $N=3$, there are infinitely many $L$ for which it is just $2$, and also infinitely many $L$ (powers of $3$) for which it is $\sim 4L$. Here, the GSD for both $N=2,3$ was calculated using techniques from commutative algebra explained in Appendix \ref{app:3dZpanisolap-gsd}. In Section \ref{sec:3dZNanisolap-N2}, we give a simpler derivation of the GSD for $N=2$ by relating the 3+1d anisotropic $\mathbb Z_2$ Laplacian model to the 3+1d anisotropic $\mathbb Z_2$ lineon model \cite{Shirley:2018nhn,Gorantla:2020jpy} on a tilted lattice.}\label{fig:intro-gsd}
\end{figure}

The $\mathbb{Z}_N$ version of the $U(1)$ Laplacian gauge theory has a large GSD and fractons, but the GSD is not robust against perturbations by local operators.
This motivates us to consider a certain anisotropic generalization, which we call the \textit{anisotropic $\mathbb{Z}_N$ Laplacian  model}.\footnote{The relation between the $\mathbb{Z}_N$ Laplacian model and its anisotropic uplift is analogous to that between the 2+1d $\mathbb{Z}_N$ Ising plaquette model \cite{paper1} and the 3+1d anisotropic lineon model in \cite{Shirley:2018nhn,Gorantla:2020jpy}.} It   has  the following salient features, some of which are reminiscent of the celebrated Haah's code \cite{Haah:2011drr}:
\begin{itemize}
\item It can be placed on a spatial lattice of the form $\Gamma\times C_{L_z}$, where $\Gamma$ is a general graph and $C_{L_z}$ is a cycle graph on $L_z$ vertices, or a 1d periodic chain with $L_z$ sites. See Figure \ref{fig:ZNanisolap-lat}.
\item The GSD is robust\footnote{On a general graph, there is no notion of locality and therefore we cannot discuss local operators and match them between the UV and the IR theories.  Consequently, the discussion of robustness is ambiguous. This is not the case on regular lattices where the usual discussion of local operators and robustness applies.  In that case, the anisotropic model is robust as we will show in Section \ref{sec:ZNanisolap-robust}.} and  is given by
\ie\label{GSDJac}
\text{GSD}=|\Jac(\Gamma,N)|^2\,,
\fe
where $\Jac(\Gamma,N)$ is a ``mod $N$-reduction'' of the Jacobian group $\Jac(\Gamma)$ of $\Gamma$.
\item It has lineons that can only move in the $z$-direction if $\Gamma$ is an infinite two-dimensional square lattice.
\item In the special case when the spatial lattice is a $L_x\times L_y\times L_z$ cubic lattice and when $N = p$ is prime, we have
\ie\label{GSDcubic}
\log_p \text{GSD} = 2\dim_{\mathbb{Z}_p} { \mathbb{Z}_p[X,Y] \over \left( \, Y(X-1)^2 +X(Y-1)^2 , X^{L_x}-1, Y^{L_y}-1 \, \right) } \,.
\fe
The definition and the explicit evaluation of this formula are discussed in Appendix \ref{app:3dZpanisolap}.
It depends on the number-theoretic properties of $L_x,L_y$. Interestingly, there exists a sequence of $L_x,L_y$ going to infinity such that the $\log_p \text{GSD} \sim O(L_x,L_y)$, but there is also a sequence such that $\log_p \text{GSD}$ stays at order 1 if $p>2$. See Figure \ref{fig:intro-gsd}.
\end{itemize}
We present this model both in terms of the low-energy limit of a stabilizer code in the Hamiltonian formalism, and in terms of a Euclidean lattice model using an integer $BF$ action \cite{Gorantla:2021svj}.
We compare the four Laplacian lattice models in Table \ref{tbl:comparison}.\footnote{In the table we assume the $\theta$-angle of the $U(1)$ Laplacian theory is not $\pi$, otherwise the GSD is 2.}

\renewcommand{\arraystretch}{2}
\begin{table}[t]
\begin{center}
\begin{tabular}{|c|c|c|c|c|}
\hline
Model &Spatial Lattice & GSD& Defects & Robust?
\tabularnewline
\hline
\hline
\makecell{Laplacian $\phi$-theory\\\cite{Gorantla:2022mrp,Gorantla:2022ssr}} & $\Gamma$ & $|\Jac(\Gamma)|$&  None & No
\tabularnewline
\hline
\makecell{$U(1)$ Laplacian gauge theory\\\cite{Gorantla:2022mrp,Gorantla:2022ssr}} & $\Gamma$ &1& Fracton &Yes
\tabularnewline
\hline
\makecell{$\mathbb{Z}_N$ Laplacian model\\Appendix \ref{sec:ZNlap}} & $\Gamma$ & $|\Jac(\Gamma,N)|$& Fracton &No
\tabularnewline
\hline
\makecell{Anisotropic $\mathbb{Z}_N$ Laplacian model\\Sections \ref{sec:ZNanisolap} and \ref{sec:3dZNanisolap}} &  $\Gamma\times C_{L_z}$ & $|\Jac(\Gamma,N)|^2$& Lineon&Yes
\tabularnewline
\hline
\end{tabular}
\end{center}
\caption{The comparison of four exotic lattice models that can be defined on a general graph $\Gamma$. The model is robust if it has no relevant local operator.  See Section \ref{sec:ZNanisolap-robust} for more details on what we mean by robustness.}\label{tbl:comparison}
\end{table}
\renewcommand{\arraystretch}{1}

Following \cite{Seiberg:2019vrp,paper1,paper2,paper3,Gorantla:2020xap,Gorantla:2020jpy,Rudelius:2020kta,Gorantla:2021svj,Gorantla:2021bda,
Burnell:2021reh,Gorantla:2022eem,Gorantla:2022mrp,Gorantla:2022ssr}, we focus on the exotic global symmetries of this model. The symmetries of the models on $\Gamma$ (the first three models in Table \ref{tbl:comparison}) are not subsystem global symmetries.  The symmetry operators are supported on most (or all) of the sites of $\Gamma$, rather than on a small subset of them.  The precise subset   depends delicately on the details of $\Gamma$.  Yet, there are many such symmetries.  In this sense these symmetries are generalizations of the dipole symmetries on cubic lattices, which are also supported on the entire lattice.  The difference is that the dipole symmetries have simple dependence on the coordinates, while here the dependence on the coordinates is more complicated.

This is not the case in the anisotropic $\mathbb{Z}_N$ Laplacian model (the fourth model in Table \ref{tbl:comparison}).  Here the symmetries act at fixed $z$ and in that sense they are subsystem symmetries.  In fact, as we will discuss below, at low energy these symmetries are independent of $z$ and are similar to one-form global symmetries.\footnote{More generally, we can classify symmetries by the difference operators that annihilate the transformation parameters $\alpha$.  For example, on a regular lattice, an ordinary symmetry has $\Delta_x\alpha=\Delta_y\alpha=\Delta_z\alpha=0$, a dipole symmetry has $\Delta_x \Delta_x\alpha=\Delta_x \Delta_y\alpha=\Delta_y \Delta_y\alpha=0$, etc.  (And of course, $\alpha $ can carry more indices for the various fields or directions in spacetime.). More interesting examples arise in theories associated to Haah's code \cite{Haah:2011drr}, where the difference equations are $(\Delta_x+\Delta_y+\Delta_z) \alpha = 0$ and $[\Delta_x \Delta_y + \Delta_y \Delta_z + \Delta_z \Delta_x + 2(\Delta_x+\Delta_y+\Delta_z)] \alpha = 0$.}

The rest of the paper is organized as follows.
In Section \ref{sec:graph}, we introduce some necessary graph theory background, including the discrete Laplace operator and the Jacobian group of  a graph.
In Section \ref{sec:ZNanisolap}, we introduce the stabilizer code and the Euclidean integer $BF$ action for the anisotropic $\mathbb{Z}_N$ Laplacian model. We derive the general expression for the GSD \eqref{GSDJac} and discuss the restricted mobility of the lineon defects from time-like symmetries.\footnote{See \cite{Gorantla:2022eem} for a definition of space-like and time-like global symmetries and their applications to the ground state degeneracy and restricted mobility constraints.}
 Section \ref{sec:3dZNanisolap} considers the special case when the spatial lattice is a cubic lattice and when $N$ is prime.
 The GSD reduces to \eqref{GSDcubic} and we discuss its asymptotic behaviors.
 Appendix \ref{app:ZNfunc} discusses  $\mathbb{Z}_N$-valued discrete harmonic functions on a general graph.
  Appendices \ref{app:3dZ2anisolap} and \ref{app:3dZpanisolap} contain the detailed computation of the GSD and the mobility restrictions for the anisotropic $\mathbb{Z}_N$ Laplacian model on a cubic lattice with $N$ a prime number.
   In Appendix \ref{sec:ZNlap}, we study the $\mathbb{Z}_N$ Laplacian model of fractons, which is not robust.

\textit{Notes added:} As we were finalizing this paper, \cite{Ebisu:2022nln} appeared on arXiv, which studies the same anisotropic $\mathbb{Z}_N$ Laplacian model  using its stabilizer code.

\section{Graph theory primer}\label{sec:graph}
In this section, we review some well-known facts about a finite graph, and $\mathbb Z_N$-valued functions on the graph. A good reference on this subject is \cite{Chung:97}. See also \cite{Gorantla:2022mrp} for more discussion on these topics in related lattice models.

Let $\Gamma$ be a simple, undirected, connected graph on $\mathsf N$ vertices.\footnote{Note that $N$ in $\mathbb Z_N$ is different from $\mathsf N$, the number of vertices of $\Gamma$.} Here, \emph{simple} means there is at most one edge between any two vertices and no self-loop on any vertex, \emph{undirected} means the edges do not have any orientation, and \emph{connected} means there is a path between any two vertices of the graph. We use $i$ to denote a vertex (or site), and $\langle i,j\rangle$ to denote an edge (or link) of the graph. We write $\langle i,j\rangle \in \Gamma$ if there is an edge between vertices $i$ and $j$ in $\Gamma$.

Let $d_i$ be the \emph{degree} of vertex $i$, i.e., the number of edges incident on $i$. The \emph{Laplacian matrix} $L$ of $\Gamma$ is an $\mathsf N \times \mathsf N$ symmetric matrix defined as follows: $L_{ii} = d_i$ for every vertex $i$, $L_{ij} = -1$ if there is an edge $\langle i,j\rangle$ between vertices $i$ and $j$, and $L_{ij} = 0$ otherwise.

\subsection{Discrete Laplacian operator $\Delta_L$ and its Smith decomposition}\label{sec:graph-disclap}
Consider a $\mathbb Z_N$-valued function $f(i)$ on the vertices of the graph. We define the \emph{discrete Laplacian} operator $\Delta_L$ as
\ie
\Delta_L f(i) := \sum_j L_{ij} f(j) = d_i f(i) - \sum_{j:\langle i,j\rangle\in \Gamma} f(j) = \sum_{j:\langle i,j\rangle\in \Gamma} [f(i) - f(j)]~,
\fe
where the equalities are modulo $N$. This is one of the most natural and universal difference operators that can be defined on any such graph $\Gamma$.

We are interested in the following two questions:
\begin{enumerate}
\item What are all the $\mathbb Z_N$-valued functions $h(i)$ that satisfy the \emph{discrete Laplacian equation}
\ie\label{graph-disclap}
\Delta_L h(i) = 0 \mod N~?
\fe
They are known as the $\mathbb Z_N$-valued \emph{discrete harmonic functions}, and we denote the set of such functions as $\mathcal H(\Gamma,\mathbb Z_N)$.\footnote{It is also known as the \emph{group of balanced vertex weightings} \cite{Berman:1986}.}
\item We define an equivalence class of $\mathbb Z_N$-valued functions by saying that two  functions $g(i)$ and $\tilde g(i)$ belong to the same class if there is a $\mathbb Z_N$-valued function $f(i)$ such that
\ie\label{graph-equiv}
\tilde g(i) - g(i) = \Delta_L f(i) \mod N~.
\fe
In this case, we write $\tilde g(i) \sim g(i)$. What are all the distinct equivalence classes under the equivalence relation ``$\sim$''?
\end{enumerate}
Interestingly, both questions can be answered using the \emph{Smith decomposition} \cite{Smith:1861} of the Laplacian matrix $L$ \cite{Lorenzini:08}. The \emph{Smith normal form} of $L$ is given by three matrices $R$, $P$, and $Q$, such that
\ie\label{graph-snf}
R = P L Q~,\qquad \text{or}\qquad R_{ab} = \sum_{i,j} P_{ai} L_{ij} Q_{jb}~,
\fe
where $P,Q\in GL_{\mathsf N}(\mathbb Z)$, and $R = \diag(r_1,\ldots,r_{\mathsf N})$. Here, $r_a$'s are nonnegative integers, known as the \emph{invariant factors} of $L$, such that $r_a$ divides $r_{a+1}$ for $a=1,\ldots, \mathsf N-1$. While $R$ is uniquely determined by $L$, the matrices $P$ and $Q$ are not. For a connected graph $\Gamma$, we have $r_a>0$ for $a=1,\ldots, \mathsf N-1$, and $r_{\mathsf N}=0$.

We state the answers to the two questions here (see Appendix \ref{app:ZNfunc} for details):
\begin{enumerate}
\item Any $\mathbb Z_N$-valued discrete harmonic function takes the form
\ie\label{graph-dischar}
h(i) = \sum_{a=1}^\mathsf{N} \frac{NQ_{ia} p_a}{\gcd(N,r_a)} \mod N~,
\fe
where $p_a = 0,\ldots, \gcd(N,r_a)-1$ for $a=1,\ldots,\mathsf N$. In other words, $\mathcal H(\Gamma,\mathbb Z_N)$ is isomorphic to the finite Abelian group $\prod_{a=1}^\mathsf{N} \mathbb Z_{\gcd(N,r_a)}$. Here, the group operation is simply the sum.  It is well-defined because if $h_1,h_2\in\mathcal H(\Gamma,\mathbb Z_N)$, then $h_1 + h_2 \in \mathcal H(\Gamma,\mathbb Z_N)$.
\item Any equivalence class is uniquely represented by the $\mathbb Z_N$-valued function
\ie\label{graph-rep}
g(i) = \sum_{a=1}^\mathsf{N} p_a (Q^{-1})_{ai} \mod N~,
\fe
where $p_a = 0,\ldots, \gcd(N,r_a)-1$ for $a=1,\ldots,\mathsf N$. In other words, the set of all equivalence classes is isomorphic to the finite Abelian group $\prod_{a=1}^\mathsf{N} \mathbb Z_{\gcd(N,r_a)}$. Here, the group operation is simply the sum which is well-defined because if $\tilde g_1 \sim g_1$ and $\tilde g_2 \sim g_2$, then $\tilde g_1 + \tilde g_2 \sim g_1 + g_2$.
\end{enumerate}
It follows that the number of $\mathbb Z_N$-valued discrete harmonic functions and the number of equivalence classes are both $\prod_{a=1}^\mathsf{N} \gcd(N,r_a)$, which is the order of the finite Abelian group $\prod_{a=1}^\mathsf{N} \mathbb Z_{\gcd(N,r_a)}$. We will have more to say about this group below.

\subsection{Jacobian group of a graph}
The finite Abelian group encountered above is intimately related to the \emph{Jacobian group} $\Jac(\Gamma)$, which is a natural finite Abelian group associated with a general graph $\Gamma$.\footnote{It has several different names in the graph theory literature, including the \emph{sandpile group} \cite{Dhar:90}, or the \emph{group of components} \cite{Lorenzini:91}, or the \emph{critical group} \cite{Biggs:1999vy} of $\Gamma$, and it is related to the \emph{group of bicycles} \cite{Berman:1986} of $\Gamma$.} In terms of the invariant factors of the Laplacian matrix $L$, we have the following isomorphism
\ie
\Jac(\Gamma) \cong \prod_{a=1}^{\mathsf N-1} \mathbb Z_{r_a}~.
\fe
The order of $\Jac(\Gamma)$ is the most fundamental and well-studied notion of \textit{complexity} in graph theory. What we have here is a ``mod $N$-reduction'' of the Jacobian group:
\ie
\Jac(\Gamma,N) \cong \prod_{a=1}^\mathsf{N} \mathbb Z_{\gcd(N,r_a)}~.
\fe
As we will see below, the group $\Jac(\Gamma,N)$ plays an crucial role in the $\mathbb Z_N$ Laplacian models.

\section{Anisotropic $\mathbb Z_N$ Laplacian model on a graph}\label{sec:ZNanisolap}
In this section, we study a robust gapped lineon model on a spatial lattice of the form $\Gamma \times C_{L_z}$, where $\Gamma$ is a simple, connected, undirected graph, and $L_z$ is the number of sites in the $z$-direction (see Figure \ref{fig:ZNanisolap-lat}). We refer to it as the anisotropic $\mathbb Z_N$ Laplacian model because it is the anisotropic extension along the $z$-direction of the $\mathbb Z_N$ Laplacian model analyzed in Appendix \ref{sec:ZNlap}.

\subsection{Hamiltonian for the stabilizer code}\label{sec:ZNanisolap-Ham}
In the Hamiltonian formulation of the anisotropic $\mathbb Z_N$ Laplacian model, there are a $\mathbb Z_N$ variable $U(i,z)$ and its conjugate variable $V(i,z)$, i.e., $U(i,z)V(i,z) = e^{2\pi i/N}V(i,z)U(i,z)$, on every site of $\Gamma\times C_{L_z}$. There are also a $\mathbb Z_N$ variable $U_z(i,z+\tfrac12)$ and its conjugate variable $V_z(i,z+\tfrac12)$, i.e., $U_z(i,z+\tfrac12)V_z(i,z+\tfrac12) = e^{2\pi i/N}V_z(i,z+\tfrac12)U_z(i,z+\tfrac12)$, on every $z$-link of $\Gamma\times C_{L_z}$.

\begin{figure}[t]
\begin{center}
~\hfill\includegraphics[scale=0.3]{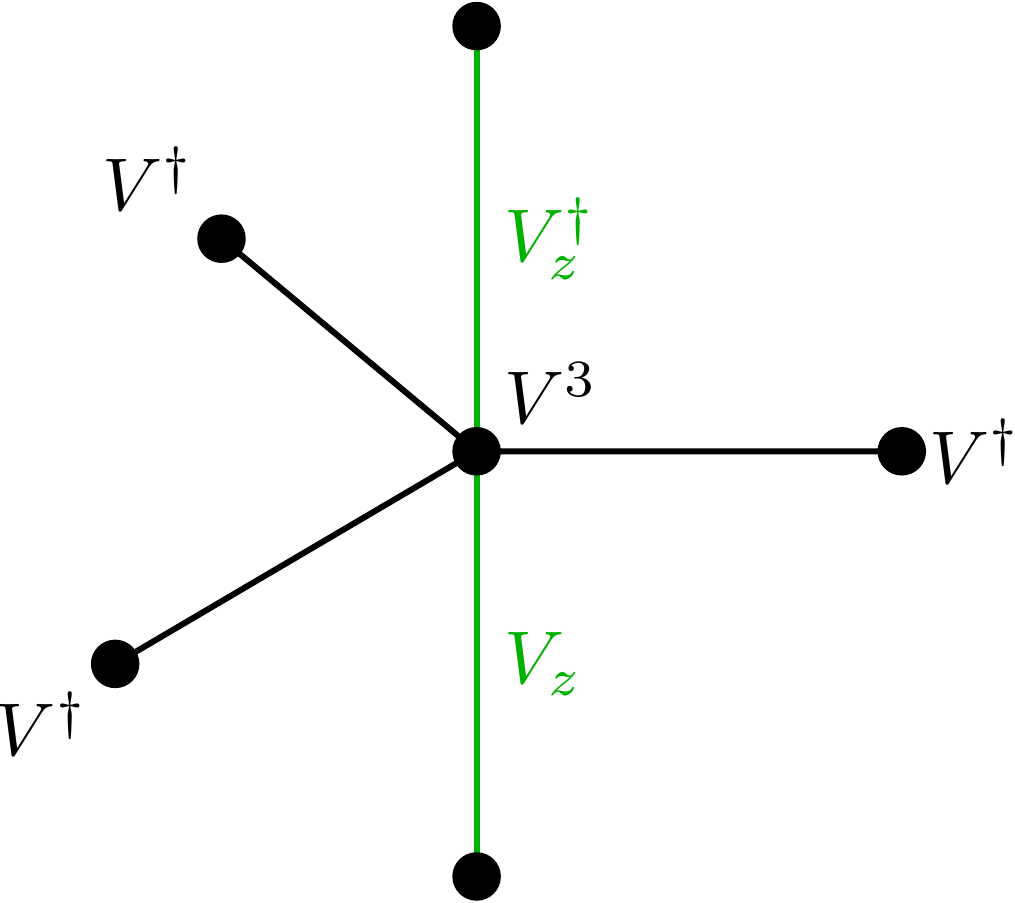}\hfill\hfill
\includegraphics[scale=0.3]{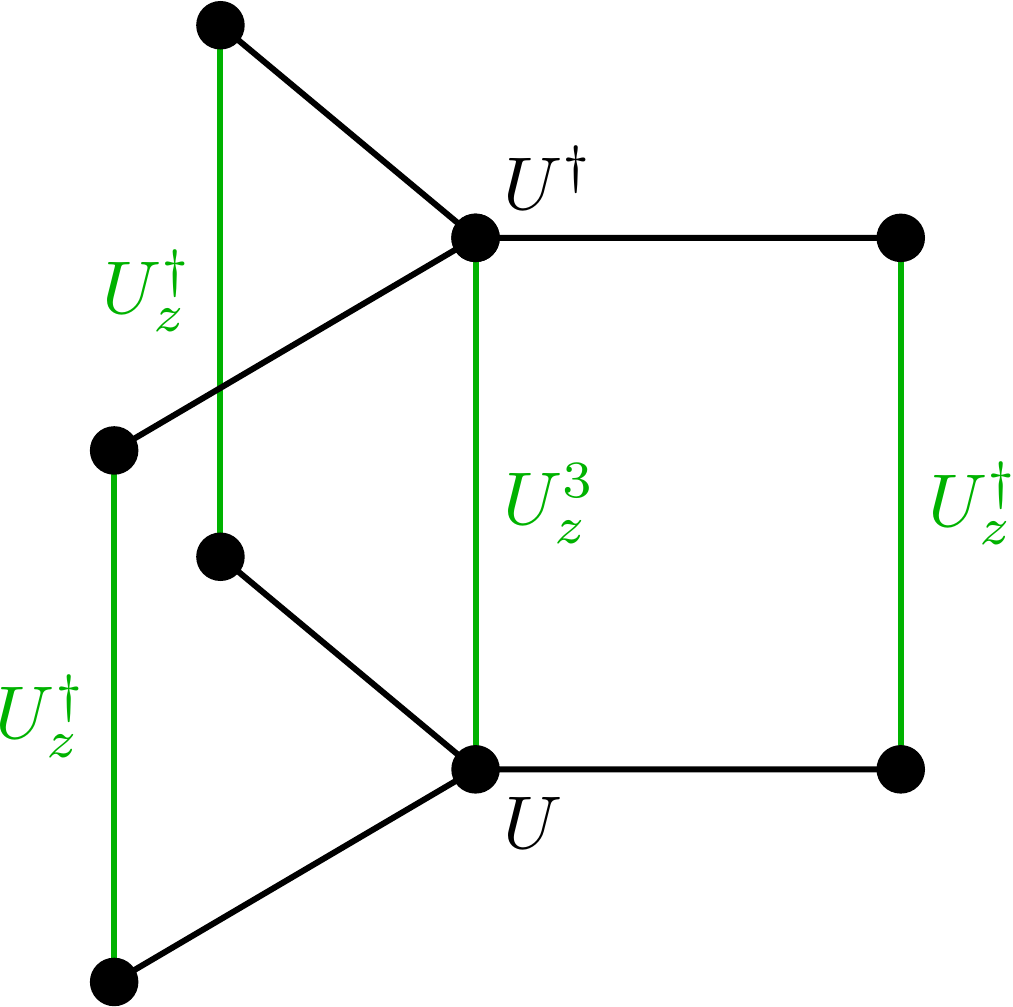}\hfill~
\\
~\hfill(a) $G$-term\hfill\hfill(b) $F$-term\hfill~
\end{center}
\caption{The two kinds of stabilizer terms in the Hamiltonian \eqref{OZNanisolap-Ham}. }\label{fig:ZNanisolap-stab}
\end{figure}

The Hamiltonian is
\ie\label{OZNanisolap-Ham}
H = -\gamma_1 \sum_{i,z} G(i,z) - \gamma_2 \sum_{i,z} F(i,z+\tfrac12) + \text{h.c.}~,
\fe
where
\ie\label{ZNanisolap-stab}
&G(i,z) = V_z(i,z+\tfrac12)^\dagger V_z(i,z-\tfrac12) \prod_{j:\langle i, j \rangle\in \Gamma} V(i,z) V(j,z)^\dagger~,
\\
&F(i,z+\tfrac12) = U(i,z+1)^\dagger U(i,z) \prod_{j:\langle i, j \rangle\in \Gamma} U_z(i,z+\tfrac12) U_z(j,z+\tfrac12)^\dagger~.
\fe
The two kinds of terms are shown in Figure \ref{fig:ZNanisolap-stab}.  Since all the terms in this Hamiltonian commute with each other, it is a stabilizer code.

The model enjoys the duality transformation
\ie
&U(i,z) \rightarrow V_z(i,z+\tfrac12)~,\qquad &&U_z(i,z+\tfrac12) \rightarrow V(i,z)~,
\\
&V(i,z) \rightarrow U_z(i,z+\tfrac12)^\dagger~,\qquad&&V_z(i,z+\tfrac12) \rightarrow U(i,z)^\dagger~.
\fe
It exchanges the two kind of terms in the Hamiltonian and therefore it maps the model with $(\gamma_1,\gamma_2)$ to the model with $(\gamma_2,\gamma_1)$.  As a result, for $\gamma_1 = \gamma_2$, this model is self-dual.

The ground states satisfy $G(i,z)=1$ and $F(i,z+\tfrac12)=1$ for all $i,z$. The excited states are violations of $G=1$ or $F=1$, which we call electric and magnetic excitations respectively. These excitations are mobile along the $z$ direction so they are at least $z$-lineons. Their mobility constraints along the graph $\Gamma$ are more complicated. We postpone that discussion to Section \ref{sec:ZNanisolap-timelikesym}.

We could also take $\gamma_1,\gamma_2 \rightarrow \infty$, in which case, the Hilbert space consists of only the ground states, and the Hamiltonian is trivial. The Euclidean presentation of this model in this limit will be discussed later in Section \ref{sec:ZNanisolap-euc}.

We are particularly interested in those operators that commute with the Hamiltonian \eqref{OZNanisolap-Ham} and act nontrivially on its ground states. They are the global symmetry operators of the model in the low energy limit, and they are also known as the logical operators of the stabilizer code. We choose a basis of these symmetry operators as follows: the electric symmetry operators are\footnote{In fact, the operator $\prod_z V(i,z)$, which is local in $\Gamma$ and extends in the $z$ direction, also commutes with the Hamiltonian. Here we choose to work with the basis of $\tilde W_z(a)$ because the latter has a simpler commutation relation with $W(a;z)$. Similarly, $\prod_z U_z(i,z+\frac12 )$ also commutes with the Hamiltonian and is local in $\Gamma$, but we choose to work in the basis of $W_z(a)$ for the same reason.}
\ie\label{ZNanisolap-ops1}
&\tilde W_z(a) = \prod_{i,z} V(i,z)^{(Q^{-1})_{ai}}~,\qquad a = 1,\ldots,\mathsf N~,
\\
&\tilde W(a;z+\tfrac12) = \prod_i V_z(i,z+\tfrac12)^{\frac{N}{\gcd(N,r_a)}Q_{ia}}~,\qquad z = 0,\ldots,L_z-1~,
\fe
and the magnetic symmetry operators are
\ie\label{ZNanisolap-ops2}
&W_z(a) = \prod_{i,z} U_z(i,z+\tfrac12)^{(Q^{-1})_{ai}}~,\qquad a = 1,\ldots,\mathsf N~,
\\
&W(a;z) = \prod_i U(i,z)^{\frac{N}{\gcd(N,r_a)}Q_{ia}}~,\qquad z = 0,\ldots,L_z-1~,
\fe
where $Q$ and $r_a$ are defined in \eqref{graph-snf}. These operators generate a $\Jac(\Gamma,N)^2$ electric symmetry and a $\Jac(\Gamma,N)^2$ magnetic symmetry.

For each $a$, the four operators in \eqref{ZNanisolap-ops1} and \eqref{ZNanisolap-ops2} are all $\mathbb Z_{\gcd(N,r_a)}$ operators. Clearly, $W(a;z)^{\gcd(N,r_a)} = \tilde W(a;z+\tfrac12)^{\gcd(N,r_a)} = 1$. (In fact, when acting on the ground states, the operators $\tilde W(a;z+\tfrac12)$ and $W(a;z)$ are independent of $z$.) Moreover, the operator $W_z(a)$ satisfies $W_z(a)^{r_a} = 1$ when acting on the ground states,\footnote{This is because $W_z(a)^{r_a} = \prod_{i,z} U_z(i,z+\tfrac12)^{r_a(Q^{-1})_{ai}} = \prod_{i,j,z} U_z(i,z+\tfrac12)^{P_{aj}L_{ji}} = \prod_j \left(\prod_z U(j,z+1) U(j,z)^\dagger\right)^{P_{aj}} = 1$, where we used the facts that $R Q^{-1} = PL$, and $F = 1$ on the ground states.} which when combined with the obvious relation $W_z(a)^N = 1$ gives the relation $W_z(a)^{\gcd(N,r_a)} = 1$. The same conclusion holds for $\tilde W_z(a)$ as well.

The basis of symmetry operators defined in \eqref{ZNanisolap-ops1} and \eqref{ZNanisolap-ops2} is chosen such that they satisfy the following commutation relations:
\ie\label{ZNanisolap-ops-comm}
W(a;z) \tilde W_z(b) =  \exp\left[ \frac{2\pi i \delta_{ab}}{\gcd(N,r_a)} \right] \tilde W_z(b) W(a;z)~,\qquad a,b = 1,\ldots,\mathsf N~,
\fe
and similarly for the other pair. So for each $a=1,\ldots,\mathsf N$, there are two independent copies of $\mathbb Z_{\gcd(N,r_a)}$ Heisenberg algebras, leading to a ground state degeneracy of\footnote{The power of 2 in \eqref{ZNanisolap-gsd} is related to the fact that the anisotropic $\mathbb Z_N$ Laplacian model is the anisotropic extension of the $\mathbb Z_N$ Laplacian model of Appendix \ref{sec:ZNlap}, whose GSD is $|\Jac(\Gamma,N)|$ \eqref{ZNlap-gsd}. This is similar to the relation $\text{GSD}_\text{3+1d anisotropic $\mathbb Z_N$ lineon model} = (\text{GSD}_\text{2+1d $\mathbb Z_N$ Ising plaquette model})^2$.\label{ftnt:power2}}
\ie\label{ZNanisolap-gsd}
\text{GSD} = \prod_{a=1}^\mathsf{N} \gcd(N,r_a)^2 = |\Jac(\Gamma,N)|^2~.
\fe

\subsection{Euclidean presentation}\label{sec:ZNanisolap-euc}
We now discuss the Euclidean presentation of the anisotropic $\mathbb Z_N$ Laplacian model. We place the theory on a Euclidean spacetime lattice $C_{L_\tau} \times \Gamma \times C_{L_z}$, where $\Gamma \times C_{L_z}$ is the spatial slice. We use $(\tau,i,z)$ to label a site in the spacetime lattice, where $i$ denotes a vertex of the graph $\Gamma$.

We use the integer $BF$ formulation of \cite{Gorantla:2021svj}. The integer $BF$-action of the anisotropic $\mathbb Z_N$ Laplacian model is
\ie\label{ZNanisolap-modVill-action}
S &= \frac{2\pi i}{N} \sum_{\tau,i,z} \Big(-\tilde m_\tau (\tau,i,z+\tfrac12) \left[ \Delta_z m(\tau,i,z+\tfrac12) - \Delta_L m_z(\tau,i,z+\tfrac12) \right]
\\
& \qquad \qquad + \tilde m_z (\tau+\tfrac12,i,z) \left[ \Delta_\tau m(\tau+\tfrac12,i,z) - \Delta_L m_\tau(\tau+\tfrac12,i,z) \right]
\\
& \qquad \qquad + \tilde m(\tau+\tfrac12,i,z+\tfrac12) \left[ \Delta_\tau m_z(\tau+\tfrac12,i,z+\tfrac12) - \Delta_z m_\tau(\tau+\tfrac12,i,z+\tfrac12) \right] \Big)~,
\fe
where the integer fields $(m_\tau,m,m_z)$ have a gauge symmetry
\ie\label{ZNanisolap-gaugesym}
&m_\tau \sim m_\tau + \Delta_\tau k + N q_\tau~,
\\
&m \sim m + \Delta_L k + N q~,
\\
&m_z \sim m_z + \Delta_z k + N q_z~,
\fe
where $k$ and $(q_\tau,q,q_z)$ are integers, and similarly for $(\tilde m_\tau,\tilde m,\tilde m_z)$. (Note that, when working modulo $N$, the second line of \eqref{ZNanisolap-gaugesym} is exactly the equivalence relation discussed in \eqref{graph-equiv}.)

The theory is self-dual under the map $(m_\tau,m,m_z) \rightarrow (\tilde m_\tau,\tilde m,\tilde m_z)$ and $(\tilde m_\tau,\tilde m,\tilde m_z) \rightarrow -(m_\tau,m,m_z)$.

The integer $BF$-action \eqref{ZNanisolap-modVill-action} describes the ground states of a stabilizer code given by the Hamitonian \eqref{OZNanisolap-Ham}. Here, we will not elaborate on the relation between the Euclidean and Hamiltonian presentations. We refer the readers to Appendix C.2 of \cite{Gorantla:2021svj} for an analogous discussion of the relation between the 2+1d $\mathbb Z_N$ toric code and the 2+1d $\mathbb Z_N$ gauge theory in the integer $BF$ presentation.

\subsubsection{Ground state degeneracy}\label{sec:ZNanisolap-gsd}

We can count the number of ground states by counting the number of solutions to the ``equations of motion'' of $(\tilde m_\tau,\tilde m,\tilde m_z)$ modulo gauge transformations:
\ie\label{ZNanisolap-eom}
&\Delta_z m - \Delta_L m_z = 0 \mod N~,
\\
&\Delta_\tau m - \Delta_L m_\tau = 0 \mod N~,
\\
&\Delta_\tau m_z - \Delta_z m_\tau = 0 \mod N~.
\fe
A gauge field $(m_\tau,m,m_z)$ that satisfies \eqref{ZNanisolap-eom} is a flat $\mathbb Z_N$ gauge field. We can use the gauge freedom in $k$ to set $m_\tau(\tau+\tfrac12,i,z)|_{\tau\ne0} = 0 \mod N$, and $m_z(\tau,i,z+\tfrac12)|_{z\ne0} = 0 \mod N$.
In this gauge choice, the last line of \eqref{ZNanisolap-eom} implies that
\ie
&\Delta_\tau m_z(\tau+\tfrac12,i,z+\tfrac12)|_{z=0} = 0 \mod N~,
\\
&\Delta_z m_\tau(\tau+\tfrac12,i,z+\tfrac12)|_{\tau=0} = 0 \mod N~.
\fe
The first two lines of \eqref{ZNanisolap-eom} then imply that
\ie
&\Delta_\tau m(\tau+\tfrac12,i,z) = 0\mod N~,
\\
&\Delta_z m(\tau,i,z+\tfrac12) = 0 \mod N~,
\fe
which in turn imply that
\ie\label{ZNanisolap-mzmt}
&\Delta_L m_z(i,z+\tfrac12)|_{z=0} = 0\mod N~,
\\
&\Delta_L m_\tau(\tau+\tfrac12,i)|_{\tau=0} = 0\mod N~.
\fe

The remaining $\tau$ and $z$-independent gauge freedom, $m(i) \sim m(i) + \Delta_L k(i)$, is exactly the equivalence relation in \eqref{graph-equiv}. So, after gauge fixing, we can set $m(i)$ to be of the form \eqref{graph-rep}, i.e., there are $|\Jac(\Gamma,N)|$ independent holonomies in $m(i)$. Since $m_z(i,z+\tfrac12)|_{z=0}$ and $m_\tau(\tau+\tfrac12,i)|_{\tau=0}$ satisfy \eqref{ZNanisolap-mzmt}, which is exactly the discrete Laplace equation \eqref{graph-disclap}, they are of the form \eqref{graph-dischar}. So there are $|\Jac(\Gamma,N)|$ independent holonomies in both of them. Finally, the set of gauge transformations $k(\tau,i,z)$ that do not act on $(m_\tau,m,m_z)$ satisfy $\Delta_\tau k = \Delta_z k = \Delta_L k = 0 \mod N$. In other words, $k(\tau,i,z)=k(i)$ is independent of $\tau,z$, and $k(i)$ satisfies the discrete Laplace equation \eqref{graph-disclap}. So such gauge transformations are of the form \eqref{graph-dischar}, and there are $|\Jac(\Gamma,N)|$ of them. Therefore, the ground state degeneracy is
\ie
\text{GSD} = \frac{|\Jac(\Gamma,N)|^3}{|\Jac(\Gamma,N)|} = |\Jac(\Gamma,N)|^2 = \prod_{a=1}^\mathsf{N}\gcd(N,r_a)^2~.
\fe

\subsubsection{Global symmetry}
There is an electric symmetry associated with the shift of $(m_\tau,m,m_z)$ by a flat $\mathbb Z_N$ gauge field. By the analysis following \eqref{ZNanisolap-eom}, up to gauge transformations, the electric (space-like) symmetry acts as\footnote{These are symmetries of the action \eqref{ZNanisolap-modVill-action} because $\Delta_\tau m$, $\Delta_z m$, and $\Delta_\tau m_z$ are clearly unaffected by the shifts, whereas $\Delta_L m_z$ is shifted by $\delta_{z,0} \Delta_L \lambda_z(i) = 0 \mod N$ because $\lambda_z(i)$ is a $\mathbb Z_N$-valued discrete harmonic function \eqref{graph-dischar}.}
\ie
&m(\tau,i,z) \rightarrow m(\tau,i,z) + \lambda(i)~,\qquad &&\lambda(i) = \sum_{a=1}^\mathsf{N} p_a (Q^{-1})_{ai}~,
\\
&m_z(\tau,i,z+\tfrac12) \rightarrow m_z(\tau,i,z+\tfrac12) + \delta_{z,0} \lambda_z(i)~,\qquad &&\lambda_z(i) = \sum_{a=1}^\mathsf{N} \frac{NQ_{ia} p_{z,a}}{\gcd(N,r_a)}~,
\fe
where $p_a$ and $p_{z,a}$ are both integers modulo $\gcd(N,r_a)$ for $a=1,\ldots,\mathsf N$. There is also a magnetic (space-like) symmetry which acts on $\tilde m$ and $\tilde m_z$ in a similar way.

The electric (space-like) symmetry is generated by the Wilson operators of $(\tilde m_\tau,\tilde m,\tilde m_z)$:
\ie\label{ZNanisolap-opstilde}
&\tilde W_z(a) = \exp\left[ \frac{2\pi i}{N} \sum_{i,z} (Q^{-1})_{ai} \tilde m_z(\tau+\tfrac12,i,z) \right]~,
\\
&\tilde W(a;z+\tfrac12) = \exp\left[ \frac{2\pi i}{\gcd(N,r_a)} \sum_i \tilde m(\tau+\tfrac12,i,z+\tfrac12) Q_{ia} \right]~,
\fe
for $a=1,\ldots,\mathsf N$ and $z=0,\ldots, L_z-1$. The electrically charged operators are the Wilson operators of $(m_\tau,m,m_z)$, i.e., $W(a;z)$ and $W_z(a)$. Similarly, the magnetic (space-like) symmetry is generated by $W(a;z)$ and $W_z(a)$, while the magnetically charged operators are $\tilde W_z(a)$ and $\tilde W(a;z+\tfrac12)$. These are the operators in \eqref{ZNanisolap-ops1} and \eqref{ZNanisolap-ops2} in the low energy limit. The commutation relation \eqref{ZNanisolap-ops-comm} can now be understood as a mixed 't Hooft anomaly between electric and magnetic space-like symmetries.

\subsubsection{Time-like symmetry and lineons}\label{sec:ZNanisolap-timelikesym}

The integer $BF$-action has defects, which extend in the time direction, such as
\ie\label{ZNanisolap-def}
W_\tau(i,z) = \exp\left[ \frac{2\pi i}{N} \sum_{\tau} m_\tau(\tau+\tfrac12,i,z) \right]~.
\fe
This describes the world-line of an infinitely heavy particle of unit charge at position $(i,z)$. It also represents the low energy limit of an electric excitation at position $(i,z)$ in the stabilizer code \eqref{OZNanisolap-Ham}. We can deform the defect to
\ie
&\exp\left[ \frac{2\pi i}{N} \sum_{\tau<0} m_\tau(\tau+\tfrac12,i,z) \right] \exp\left[ \frac{2\pi i}{N} \sum_{z \le z'' < z'} m_z(0,i,z''+\tfrac12) \right]
\\
&\qquad \times \exp\left[ \frac{2\pi i}{N} \sum_{\tau\ge0} m_\tau(\tau+\tfrac12,i,z') \right]~.
\fe
This configuration describes a particle moving along the $z$-direction.

Next, we discuss the mobility of the particle along the graph $\Gamma$. Such a motion is constrained by the time-like global symmetry, which acts on extended defects rather than the operators or states of the Hilbert space (see \cite{Gorantla:2022eem} for more discussions on time-like global symmetries). Up to gauge transformation, the electric time-like symmetry acts as\footnote{This is a symmetry of the action \eqref{ZNanisolap-modVill-action} because $\Delta_z m_\tau$ is clearly unaffected by the shift, and $\Delta_L m_\tau$ is shifted by $\delta_{\tau,0} \Delta_L\lambda_\tau(i) = 0 \mod N$ because $\lambda_\tau(i)$ is a $\mathbb Z_N$-valued discrete harmonic function \eqref{graph-dischar}.}
\ie
&m_\tau(\tau+\tfrac12,i,z) \rightarrow m_\tau(\tau+\tfrac12,i,z) + \delta_{\tau,0} \lambda_\tau(i)~,\qquad &&\lambda_\tau(i) = \sum_{a=1}^\mathsf{N} \frac{N Q_{ia} p_{\tau,a}}{\gcd(N,r_a)}~,
\fe
where $p_{\tau,a} = 0,\ldots, \gcd(N,r_a)-1$. Hence, the group of electric time-like symmetry is $\Jac(\Gamma,N)$.

Two defects at sites $(i,z)$ and $(i',z')$ carry the same time-like charges, or equivalently, a particle can hop from $(i,z)$ to $(i',z')$, if and only if\footnote{Indeed, when this condition holds, the defect that ``moves'' a particle from $(i,z)$ to $(i',z')$ at time $\tau=0$ is given by
\ie
&\exp\left[ \frac{2\pi i}{N} \sum_{\tau<0} m_\tau(\tau+\tfrac12,i,z) \right]\exp\left[ -\frac{2\pi i}{N} \sum_{a,j} \left(\frac{Q_{ia} - Q_{i'a}}{\gcd(N,r_a)}\right) \tilde r_a P_{aj} m(0,j,z) \right]
\\
&\times \exp\left[ \frac{2\pi i}{N} \sum_{z \le z'' < z'} m_z(0,i',z''+\tfrac12) \right] \exp\left[ \frac{2\pi i}{N} \sum_{\tau\ge0} m_\tau(\tau+\tfrac12,i',z') \right]~,
\fe
where for each $a$, $\tilde r_a$ is the integer solution of the equation $\tilde r_a r_a = \gcd(N,r_a)\mod N$.}
\ie\label{ZNanisolap-mob}
Q_{ia} = Q_{i'a} \mod \gcd(N,r_a)~,\qquad \forall a=1,\ldots,\mathsf N~.
\fe
In other words, the time-like charges $Q_{ia}$ encode the superselection sector of a defect.

Similarly, there are defects of $\tilde m_\tau$ which represent the low energy limit of magnetic excitations of the stabilizer code \eqref{OZNanisolap-Ham}. By the self-duality, similar mobility restrictions apply to the defects of $\tilde m_\tau$ due to a $\Jac(\Gamma,N)$ dual magnetic time-like symmetry.

While this selection rule \eqref{ZNanisolap-mob} is not very intuitive, we will give strong mobility constraints in the special case where the spatial lattice is a cubic lattice (i.e., $\Gamma$ is a 2d torus graph $C_{L_x} \times C_{L_y}$) in Section \ref{sec:3dZNanisolap}. In particular, under some mild conditions, the particles can move only along the $z$-direction, i.e., they are lineons.

\subsubsection{Robustness}\label{sec:ZNanisolap-robust}

Let us examine the robustness of the low-energy theory. Typically, in order to address this question we should map local operators in the UV theory to local operators in the IR theory.  However, if $\Gamma$ is a general graph, it has no notion of locality and we cannot discuss local operators.  Therefore, the usual discussion of robustness does not apply.  Instead, we will restrict $\Gamma$ to be a regular lattice (such as square lattice, honeycomb lattice, cubic lattice, etc.), where there is an unambiguous notion of locality and we can consider localized operators.  (One might be able to extend the discussion to the case of an infinite graph $\Gamma$ with some restrictions on its connectivity.  We will not attempt to do it here.)

The only operators that act nontrivially on the ground states are $\tilde W(a;z)$ and $\tilde W_z(a)$ of \eqref{ZNanisolap-opstilde}, and similarly, $W(a;z)$ and $W_z(a)$. It is clear that $W_z(a)$ and $\tilde W_z(a)$ are supported over $L_z$ sites in the $z$-direction, so they are not finitely supported in the infinite volume limit. Now, we show that $W(a;z)$ is not finitely supported when $\Gamma$ is a regular lattice. Assume to the contrary that it is finitely supported. It generates a $\Jac(\Gamma,N)$ magnetic symmetry that shifts the gauge field $\tilde m(\tau+\tfrac12,i,z')$ by $\delta_{z',z} f(i)$, where $f(i)$ is a $\mathbb Z_N$-valued discrete harmonic function. The support of $f(i)$ is precisely the support of the operator $W(a;z)$, so $f(i)$ is also finitely supported. However, on a regular lattice, there is no nontrivial finitely-supported discrete harmonic function.\footnote{For example, on a square lattice with coordinates $(x,y)$, let $f(x,y)$ be a finitely-supported discrete harmonic function. Consider a large rectangular region $R$ that contains the support of $f(x,y)$, i.e., $f(x,y) = 0$ for $(x,y)$ outside $R$. Using the discrete Laplace equation $\Delta_L f(x,y)=0$ at the points immediately outside $R$, one can show that $f(x,y)=0$ at the points immediately inside $R$. By induction, $f(x,y)=0$ everywhere inside $R$. This argument extends to any regular lattice. It also extends to a more general class of graphs but we do not discuss this here.} Therefore, $W(a;z)$ cannot be finitely supported. Similarly, $\tilde W(a;z+\tfrac12)$ is also not finitely supported.

Since there are no finitely-supported operators that act nontrivially in the space of ground states, the anisotropic $\mathbb Z_N$ Laplacian model is robust.  We can deform the microscopic model with finitely-supported operators.   As long as their coefficients are small enough, they map to localized deformations of the low-energy theory.  However, since there are no local point-like operators acting in the low-energy theory, it cannot change.

\section{3+1d anisotropic $\mathbb Z_N$ Laplacian model on a torus}\label{sec:3dZNanisolap}
In this section, we analyze the GSD and restricted mobility of the anisotropic $\mathbb Z_N$ Laplacian model on an $L_x \times L_y \times L_z$ cubic lattice with periodic boundary condition, i.e., $\Gamma$ is a 2d torus graph $C_{L_x} \times C_{L_y}$. On $\Gamma=C_{L_x} \times C_{L_y}$, we have the following identification between the lattice points:
\ie\label{3dZ2anisolap-id}
(x,y) \sim (x+L_x, y) \sim (x,y+L_y)~.
\fe
Throughout this section, we use $(x,y)$ to denote a vertex of the 2d torus graph, and reserve $i$ to denote a vertex of a general graph $\Gamma$. Then, the discrete Laplacian operator $\Delta_L$ takes the more familiar form $\Delta_x^2 + \Delta_y^2$ in the $xy$-plane.

\subsection{Upper bound on $\log_N\mathrm{GSD}$}

It is clear from \eqref{ZNanisolap-gsd} that the GSD depends only on properties of $\Gamma$ and is independent of $L_z$.  Here, we give an upper bound on how fast $\log_N\text{GSD}$ can grow with $L_x,L_y$.

Recall that the GSD of the anisotropic $\mathbb Z_N$ Laplacian model is $|\Jac(\Gamma,N)|^2$ \eqref{ZNanisolap-gsd}. As we showed in Section \ref{sec:graph-disclap}, $|\Jac(\Gamma,N)|$ is also the number of equivalence classes under the equivalence relation ``$\sim$'' in \eqref{graph-equiv}. Combining these facts, we have
\ie\label{gsd-equivcl}
\text{GSD} = \left|\frac{\{g(i)\}}{g(i) \sim g(i) + \Delta_L f(i)}\right|^2~,
\fe
where $g(i)$ and $f(i)$ are $\mathbb Z_N$-valued functions on the graph $\Gamma$.

When $\Gamma$ is the 2d torus graph $C_{L_x} \times C_{L_y}$, interpreting the equivalence relation as a gauge symmetry, we can gauge fix $g(x,y) = 0 \mod N$ everywhere except at $x=0,1$, or at $y=0,1$. In other words, the number of sites where $g(x,y) \ne 0$ after gauge fixing is at most $2\min(L_x,L_y)$. Since $g(x,y)$ is $\mathbb Z_N$-valued, it follows that the number of nontrivial configurations of $g(x,y)$ is at most $N^{2\min(L_x,L_y)}$. Therefore,
\ie\label{3dZNanisolap-gsdbound}
\log_N\text{GSD} \le 4\min(L_x,L_y)~.
\fe

\subsection{$N=2$}\label{sec:3dZNanisolap-N2}
\begin{figure}[t]
\begin{center}
\includegraphics[scale=0.3]{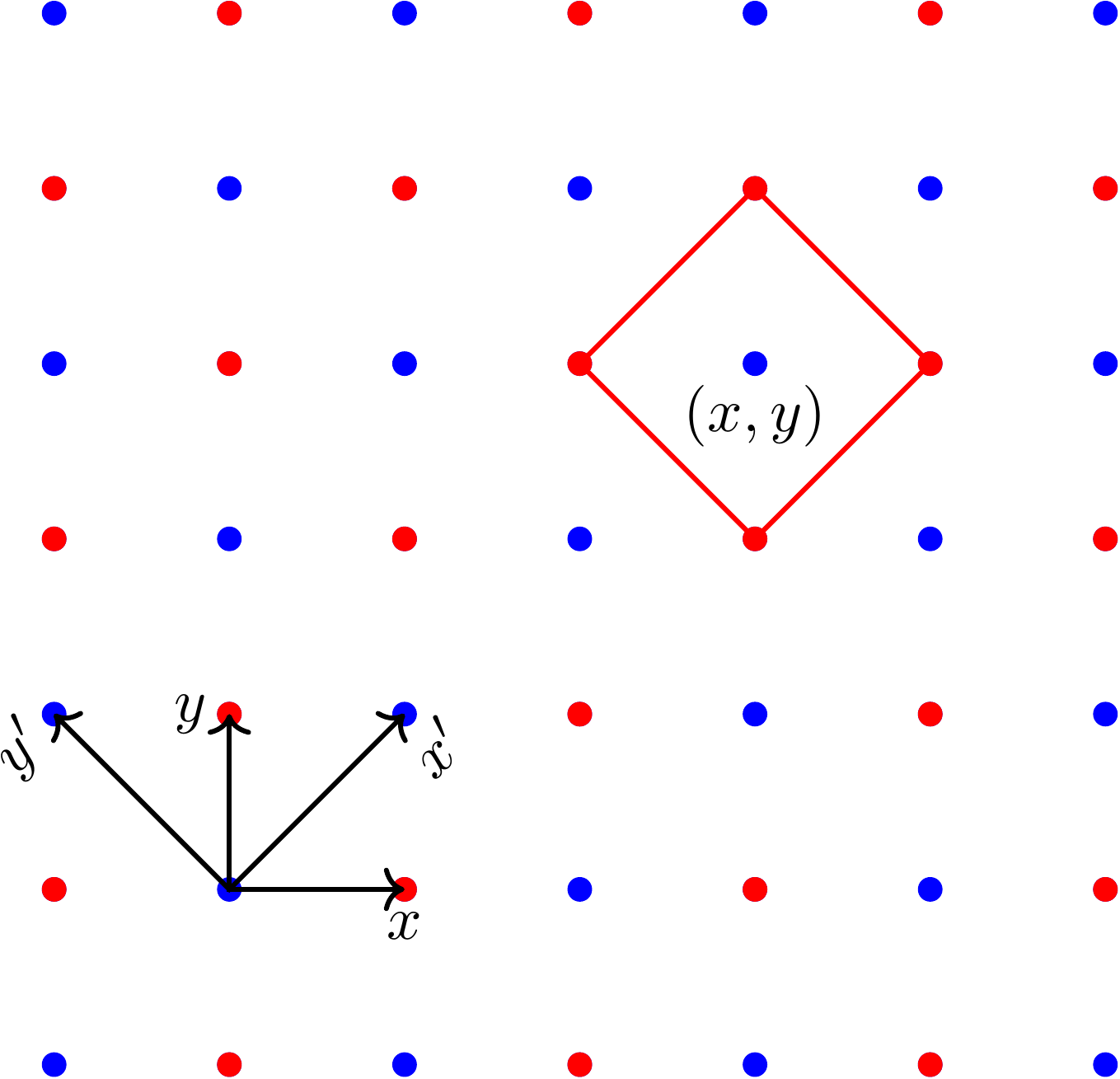}
\end{center}
\caption{The stabilizer terms \eqref{3danisolap-stab} of the 3+1d anisotropic $\mathbb Z_2$ Laplacian model are equivalent to those of the 3+1d anisotropic $\mathbb Z_2$ lineon model of \cite{Shirley:2018nhn,Gorantla:2020jpy} on a ``$\frac{\pi}{4}$-tilted'' lattice (the $z$-direction is suppressed in this figure). In particular, the product of $V$'s in $G$ and the product of $U_z$'s in $F$ involve only the four sites around the ``$\frac{\pi}{4}$-tilted plaquette'' centered at $(x,y)$. Such a plaquette consists of only red sites, or only blue sites but not both. One such red tilted plaqutte is shown. The red and blue sublattices might or might not give independent copies of the 3+1d anisotropic $\mathbb Z_2$ lineon model depending on the parities of $L_x$ and $L_y$ in the identifications \eqref{3dZ2anisolap-id}. We use the coordinates $(x',y')$ for the tilted lattice, which are related to the original coordinates $(x,y)$ as $x' = \frac{x+y}{2}$ and $y'=\frac{x-y}{2}$. They are integers on the blue sublattice, and half-integers on the red sublattice.}\label{fig:N2tilt}
\end{figure}

When $N=2$, the stabilizer terms \eqref{ZNanisolap-stab} simplify to
\ie\label{3danisolap-stab}
G(x,y,z) &= V_z(x,y,z+\tfrac12) V_z(x,y,z-\tfrac12) \prod_{\epsilon_x,\epsilon_y=\pm1} V(x+\epsilon_x,y+\epsilon_y,z)~,
\\
F(x,y,z+\tfrac12) &= U(x,y,z+1) U(x,y,z) \prod_{\epsilon_x,\epsilon_y=\pm1} U_z(x+\epsilon_x,y+\epsilon_y,z+\tfrac12)~.
\fe
Here, the product of the four $V$'s in $G$ and the product of the four $U_z$'s in $F$ both involve only the four sites around the ``$\frac{\pi}{4}$-tilted plaquette'' centered at $(x,y)$. This is illustrated in Figure \ref{fig:N2tilt}. Therefore, each stabilizer term of the 3+1d anisotropic $\mathbb Z_2$ Laplacian model is equivalent to a stabilizer term of the 3+1d anisotropic $\mathbb Z_2$ lineon model of \cite{Shirley:2018nhn,Gorantla:2020jpy} on a ``$\frac{\pi}{4}$-tilted'' lattice.

Let us define the coordinates $(x',y') = (\frac{x+y}{2},\frac{y-x}{2})$ for the tilted lattice. The tilted lattice decomposes into two sublattices: those with integral $(x',y')$ and those with half-integral $(x',y')$. These are shown in blue and red in Figure \ref{fig:N2tilt}. Observe that any tilted plaquette consists of sites from only one of the sublattices. Therefore, locally, there are two copies of the 3+1d anisotropic $\mathbb Z_2$ lineon model, one on each sublattice. On an infinite lattice, these two copies are independent. However, the identifications \eqref{3dZ2anisolap-id} can couple them: when $L_x$ and $L_y$ are both even, the two sublattices are decoupled and there are two copies of the 3+1d anisotropic $\mathbb Z_2$ lineon model, whereas when $L_x$ or $L_y$ is odd, the two sublattices are identified, so there is only one copy of the 3+1d anisotropic $\mathbb Z_2$ lineon model. In all these cases, the identifications on the tilted lattice for the 3+1d anisotropic $\mathbb Z_2$ lineon model are given in \eqref{N2ee-id}, \eqref{N2oe-id}, and \eqref{N2oo-id}.

We present the GSD and mobility restrictions in this model for $L_x = L_y = L$ and refer the readers to Appendix \ref{app:3dZ2anisolap} on results for arbitrary $L_x$ and $L_y$. The ground state degeneracy is given by
\ie\label{N=2GSD}
\text{GSD} = \begin{cases}
2^{4L}~,\qquad & L \text{ even}~,
\\
2^{4L-2}~,\qquad & L \text{ odd}~.
\end{cases}
\fe
This is in agreement with the plot for $N=2$ in Figure \ref{fig:intro-gsd}, and it saturates the bound in \eqref{3dZNanisolap-gsdbound} when $L$ is even. Furthermore, a $z$-lineon cannot hop between different sites in the $xy$-plane. In contrast, a dipole of $z$-lineons separated in the $(1,\pm1)$ direction can move in the $(1,\mp1)$ direction in the $xy$-plane. These mobility restrictions follow from the relation between the 3+1d anisotropic $\mathbb Z_2$ Laplacian model and the 3+1d anisotropic $\mathbb Z_2$ lineon model on the tilted lattice.

To conclude, the $N=2$ anisotropic Laplacian model is made out of the known anisotropic lineon model  of \cite{Shirley:2018nhn,Gorantla:2020jpy}, with a relatively simple GSD \eqref{N=2GSD}.
The next subsection discusses the anisotropic $\mathbb{Z}_p$ Laplacian model with $p$ an odd prime, which is a genuinely new model and has a much more intricate GSD.

\subsection{$N=p$ prime larger than $2$}\label{sec:3dZNanisolap-Np}

When $N=p$ is a prime larger than $2$ we can follow \cite{Haah_2013} and use techniques from commutative algebra to compute the ground state degeneracy. We show that the GSD is given by \eqref{GSDcubic}. See Appendix \ref{app:3dZpanisolap-gsd} for the meaning and derivation of this expression.

The expression in \eqref{GSDcubic} can be simplified in some special cases. Let $q\ne p$ be another odd prime such that $p$ is a primitive root modulo $q^m$, where $m\ge1$, i.e., $p$ is the generator of the multiplicative group of integers modulo $q^m$, denoted as $\mathbb Z_{q^m}^\times$.\footnote{For any positive integer $n$, the set of all integers $a$ such that $1\le a<n$ and $\gcd(a,n)=1$ form a group under multiplication, known as the multiplicative group of integers modulo $n$, and denoted as $\mathbb Z_n^\times$. It is cyclic exactly when $n = 1,2,4,q^m$, or $2q^m$, where $q$ is an odd prime and $m\ge1$ \cite[Section~2.8]{niven1991}. Whenever $\mathbb Z_n^\times$ is cyclic, it has a single generator, and the notion of ``primitive root modulo $n$'' is well-defined.} Then, for $L_x = p^{k_x} q^m$ and $L_y = p^{k_y} q^m$, where $k_x,k_y,m\ge0$, we show that
\ie
\log_p\text{GSD} = 2\left[2p^{\min(k_x,k_y)} - \delta_{k_x,k_y}\right]~.
\fe
We see that the bound \eqref{3dZNanisolap-gsdbound} is saturated whenever $k_x \ne k_y$ and $m=0$, i.e., there are infinitely many $L_x,L_y$ for which $\log_p\text{GSD}$ scales as $O(L_x,L_y)$. On the other hand, when $k_x = k_y = 0$, we have $\log_p\text{GSD}=2$ for any $m$, i.e., there are also infinitely many $L_x,L_y$ for which $\log_p\text{GSD}$ remains finite.

The last statement relies on the existence of an odd prime $q$ such that $p$ is a primitive root modulo $q^m$ for all $m\ge1$. A sufficient condition for this is that $p$ is a primitive root modulo $q^2$ \cite[Section~2.8]{niven1991}. For example, $3$ is a primitive root modulo $5^2$, so for $p=3$, we can choose $q=5$. Similarly, for $p=5,7$, we can choose $q=7,11$ respectively. In fact, one can verify numerically that for all $p\lesssim 10^9$, there is such a $q$. However, there is no proof of existence of such $q$ for arbitrary $p$.

Interestingly, \emph{Artin's conjecture on primitive roots} \cite{Moree2012} states that there are infinitely many prime $q$ such that $p$ is a primitive root modulo $q$.\footnote{Note that $p$ being a primitive root modulo $q$ does not imply that $p$ is a primitive root modulo $q^2$.} Whenever $L_x = L_y = q$ for any such $q$, we find that $\log_p\text{GSD} = 2$. This gives another infinite family of $L_x,L_y$ for which $\log_p\text{GSD}$ remains finite. However, Artin's conjecture is still unproven, except under the assumption of the \emph{generalized Riemann hypothesis} \cite{Hooley1967}, which is also unproven.

We can apply similar techniques to determine the mobility of $z$-lineons in the $xy$-plane as well. (See Appendix \ref{app:3dZpanisolap-mob} for more details.) There exist certain special values of $L_x,L_y$ (e.g., $L_x = L_y = q^m$, where $q$ is an odd prime such that $p$ is a primitive root modulo $q^m$) for which the $z$-lineons are completely mobile in the $xy$-plane. However, on an infinite square lattice, any finite set of $z$-lineons is completely immobile (unless they can be annihilated), assuming that their charges and the separations between them are fixed during the motion, i.e., they cannot move ``rigidly.''

It is surprising that the set of $L_x,L_y$ for which $\log_p\text{GSD}$ remains finite and the $z$-lineons are completely mobile is intimately related to well-known open problems in number theory.

\section*{Acknowledgements}
We are grateful to H.\ Ebisu, E.\ Studnia, and S.\ Velusamy for helpful discussions. PG was supported by the Physics Department of Princeton University. HTL is supported in part by a Croucher fellowship from the Croucher Foundation, the Packard Foundation and the Center for Theoretical Physics at MIT. The work of NS was supported in part by DOE grant DE$-$SC0009988 and by the Simons Collaboration on Ultra-Quantum Matter, which is a grant from the Simons Foundation (651440, NS). The work of SHS was supported in part by NSF grant PHY-2210182. The authors of this paper were ordered alphabetically.

\appendix

\section{More on $\mathbb Z_N$-valued functions on a graph}\label{app:ZNfunc}

In this appendix, we analyze the space of $\mathbb{Z}_N$-valued harmonic functions and the equivalence classes of $\mathbb{Z}_N$-valued functions on a general graph $\Gamma$. We use the Smith decomposition \eqref{graph-snf} of the Laplacian matrix $L$ to give complete answers of these two questions  mentioned in Section \ref{sec:graph-disclap}.

Recall that the Smith normal form of $L$ is given by $R = P L Q$, where $P,Q\in GL_{\mathsf N}(\mathbb Z)$, and $R = \diag(r_1,\ldots,r_{\mathsf N})$. Here, $r_a$'s are nonnegative integers such that $r_a$ divides $r_{a+1}$ for $a=1,\ldots, \mathsf N-1$. While $R$ is uniquely determined by $L$, the matrices $P$ and $Q$ are not.

In the index notation of \eqref{graph-snf}, we have $R_{ab} = r_a \delta_{ab} = \sum_{i,j} P_{ai}L_{ij}Q_{jb}$. While all the indices here run from $1$ to $\mathsf N$, only $i,j$ have a natural interpretation as vertices of the graph $\Gamma$.

We can now answer the first question raised in Section \ref{sec:graph-disclap}: find all the $\mathbb Z_N$-valued discrete harmonic functions. We first transform the $\mathbb Z_N$-valued function $h(i)$ to a new basis:
\ie\label{disclap-f-basis}
h'_a = \sum_i (Q^{-1})_{ai} h(i) \mod N~,
\fe
In this basis, the discrete Laplace equation \eqref{graph-disclap} is ``diagonal'':
\ie\label{disc-Poisson-ind}
r_a h'_a = 0 \mod N~,\qquad a = 1,\ldots,\mathsf N~.
\fe
We can solve this equation independently for each $a$. The most general solution is
\ie
h'_a = \frac{N p_a}{\gcd(N,r_a)} \mod N~,\qquad a = 1,\ldots,\mathsf N~,
\fe
where $p_a = 0,\ldots,\gcd(N,r_a)-1$. Transforming back to the original basis, the most general $\mathbb Z_N$-valued discrete harmonic function is
\ie
h(i) = \sum_{a=1}^\mathsf{N} \frac{NQ_{ia} p_a}{\gcd(N,r_a)} \mod N~.
\fe

Let us now address the second question raised in Section \ref{sec:graph-disclap}: find all the equivalence classes under the equivalence relation ``$\sim$''. Since the Laplacian matrix $L$ is symmetric, taking the transpose of $R=PLQ$ gives another Smith decomposition $R = Q^T L P^T$. Using this, we transform the $\mathbb Z_N$-valued function $g(i)$ to a (different) new basis
\ie
g''_a = \sum_i g(i) Q_{ia} \mod N~.
\fe
We define $\tilde g''_a$ similarly for another function $\tilde g(i)$ in the same equivalence class. In this basis, the equivalence relation \eqref{graph-equiv} is ``diagonal'':
\ie
\tilde g''_a - g''_a = r_a \hat f_a \mod N~,
\fe
where $\hat f_a = \sum_i f(i) P^{-1}_{ia} \mod N$. Therefore, the equivalence class of $g(i)$ is completely determined by $\mathsf N$ congruence classes:
\ie
g''_a \mod \gcd(N,r_a)\,,~~a=1,\ldots,\mathsf N\,.
\fe
Going back to the original basis, a representative of the equivalence class ``$p_a \mod \gcd(N,r_a)$'' is
\ie
g(i) = \sum_{a=1}^\mathsf{N} p_a (Q^{-1})_{ai} \mod N~,
\fe
where $p_a = 0,\ldots,\gcd(N,r_a)-1$ for $a = 1,\ldots,\mathsf N$.

\section{3+1d anisotropic $\mathbb Z_2$ Laplacian model}\label{app:3dZ2anisolap}
In this appendix, we use the relation between the 3+1d anisotropic $\mathbb Z_2$ Laplacian model of Section \ref{sec:3dZNanisolap-N2} and the 3+1d anisotropic $\mathbb Z_2$ lineon model to compute the GSD and restricted mobility of the former.

\subsection{Ground state degeneracy}
Recall that the stabilizer terms of the 3+1d anisotropic $\mathbb Z_2$ Laplacian model are given by \eqref{3danisolap-stab}, which are equivalent to those of the 3+1d anisotropic $\mathbb Z_2$ lineon model of \cite{Shirley:2018nhn,Gorantla:2020jpy} on a tilted lattice. Moreover, the latter is an anisotropic extension of the 2+1d $\mathbb Z_2$ Ising plaquette model \cite{paper1} on the tilted lattice.

Now, the identifications on the original lattice are \eqref{3dZ2anisolap-id}
\ie
(x,y) \sim (x+L_x, y) \sim (x,y+L_y)~,
\fe
In the new coordinates $(x',y') = (\frac{x+y}{2},\frac{y-x}{2})$, the identifications on the tilted lattice take the schematic form
\ie
(x',y') \sim (x'+L^u_{x'},y'+L^u_{y'}) \sim (x'+L^v_{x'},y'+L^v_{y'})~.
\fe
The authors of \cite{Rudelius:2020kta} analyzed the 2+1d $\mathbb Z_2$ Ising plaquette model on a 2d spatial torus with such identifications. Their strategy was to reduce the identifications to the form
\ie\label{minimal-reduced-id}
(x',y') \sim (x'+ML^\text{eff}_{x'},y') \sim (x'+KL^\text{eff}_{x'},y'+L^\text{eff}_{y'})~,
\fe
where $\gcd(M,K) = 1$. Then, they showed that
\ie
\text{GSD}_\text{2+1d $\mathbb Z_2$ Ising plaq} = \gcd(2,M) \cdot 2^{L^\text{eff}_{x'} + L^\text{eff}_{y'} - 1}~.
\fe
It follows that (see footnote \ref{ftnt:power2})
\ie
\text{GSD}_\text{3+1d aniso $\mathbb Z_2$ lineon} = \left[\gcd(2,M) \cdot 2^{L^\text{eff}_{x'} + L^\text{eff}_{y'} - 1}\right]^2~.
\fe

\begin{figure}[t]
\begin{center}
~\hfill\includegraphics[scale=0.2]{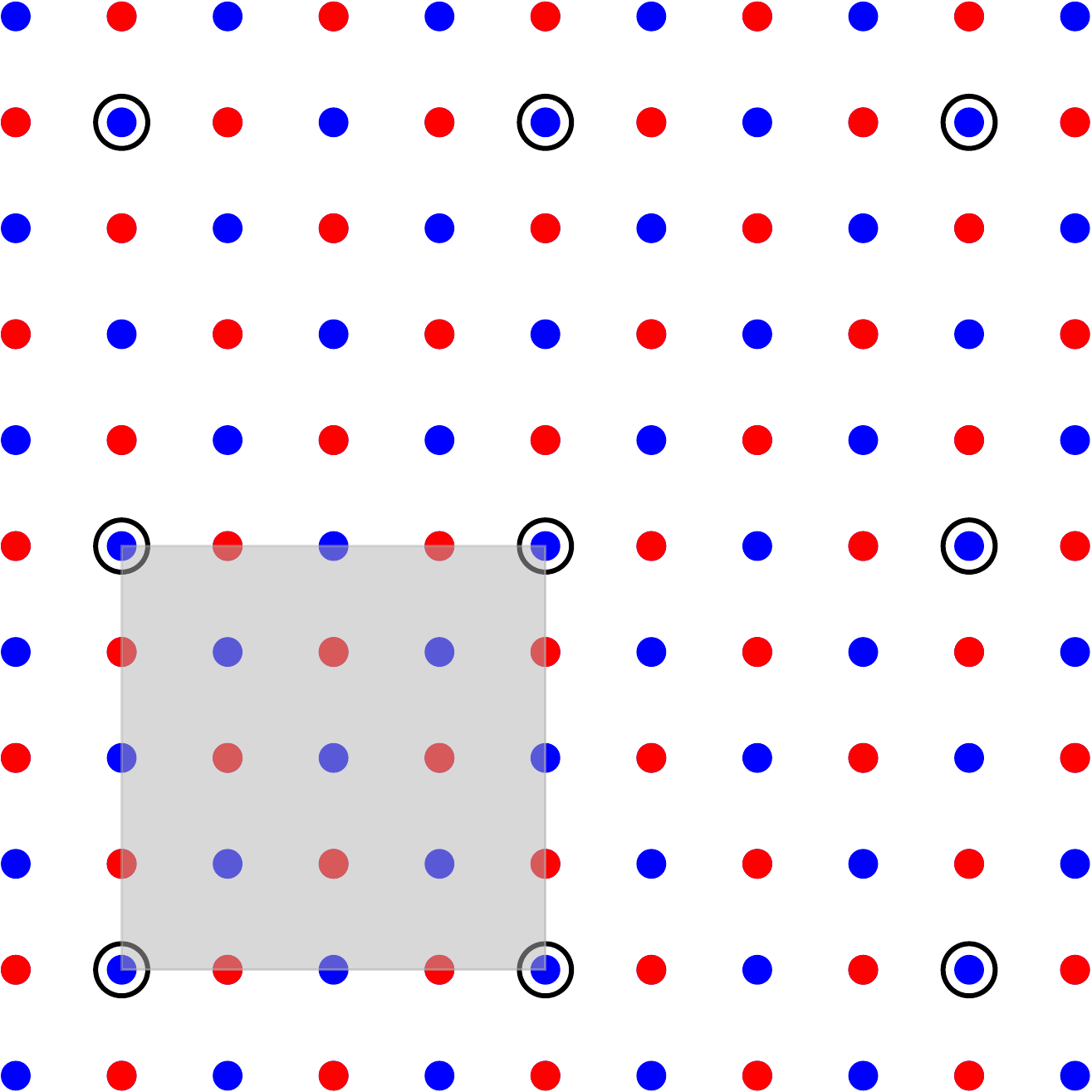}\hfill\hfill
\includegraphics[scale=0.2]{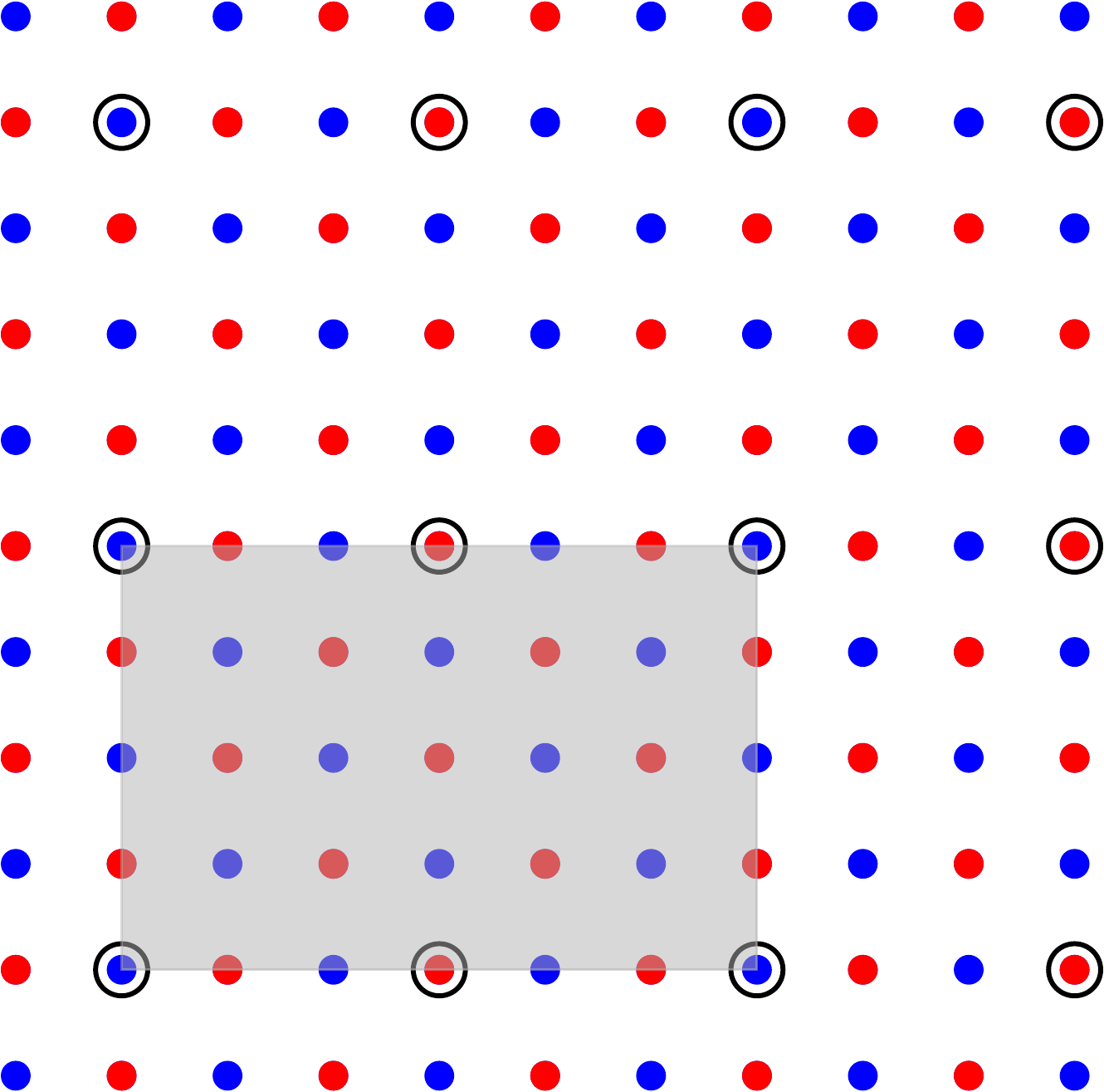}\hfill\hfill
\includegraphics[scale=0.2]{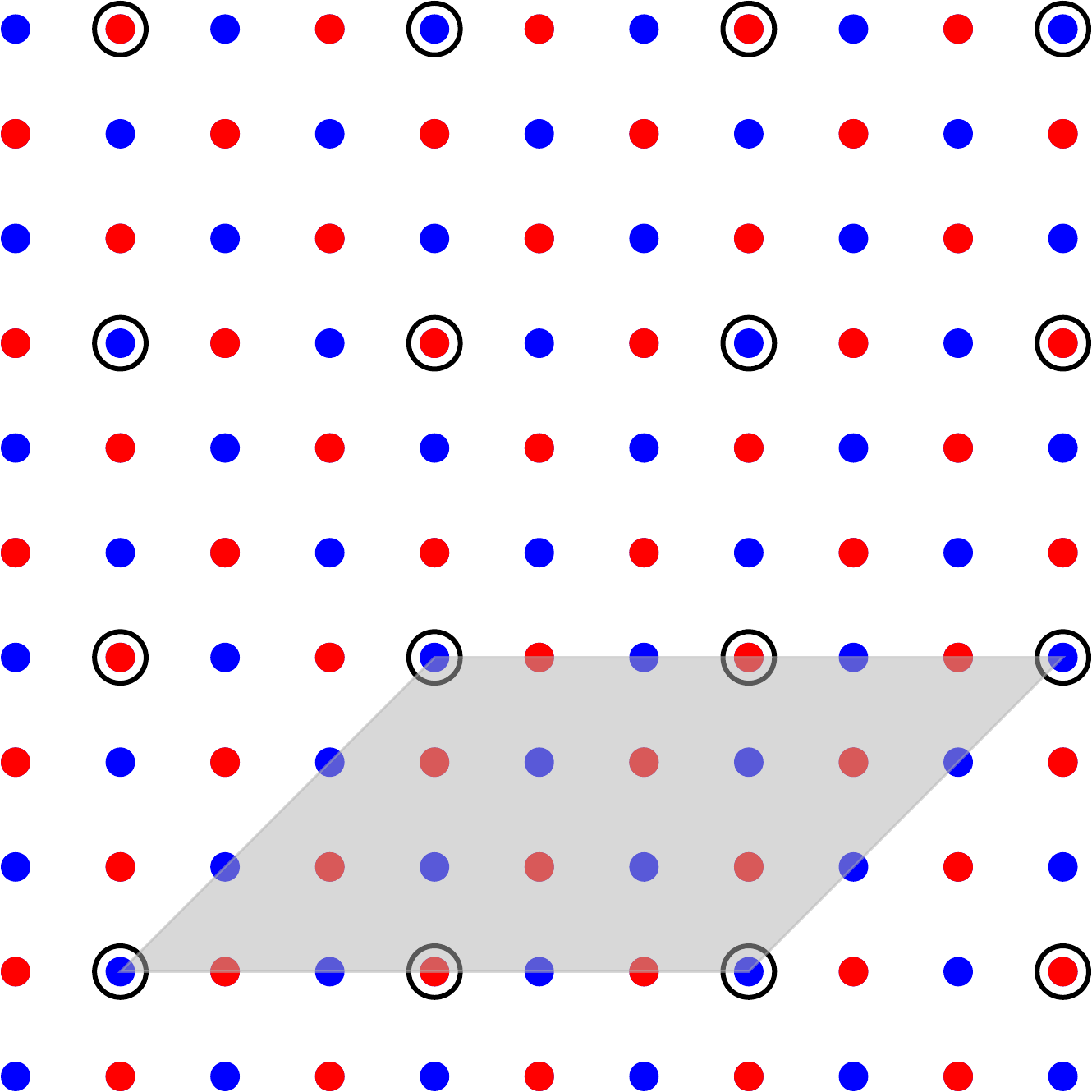}\hfill~
\\
~\hfill (a) $L_x,L_y$ even \hfill \hfill (b) $L_x$ odd, $L_y$ even\hfill \hfill (c) $L_x,L_y$ odd ~~\hfill~
\end{center}
\caption{The minimal identifications on the blue sublattice of Figure \ref{fig:N2tilt} for different parities of $L_x$ and $L_y$. The black circles represent the identifications \eqref{3dZ2anisolap-id}, and the shaded regions represent the fundamental domains of the blue sublattice under these identifications.}\label{fig:N2parity}
\end{figure}

Let us use these results to compute the GSD of the 3+1d anisotropic $\mathbb Z_2$ Laplacian model:

\begin{itemize}
\item \underline{$L_x,L_y$ even}: In this case, the identifications in \eqref{3dZ2anisolap-id} do not couple the two sublattices. So, there are two independent copies of the the 3+1d anisotropic $\mathbb Z_2$ lineon model, and we can work with one copy at a time, say the blue sublattice in Figure \ref{fig:N2tilt}. In the new coordinates $(x',y')$, the minimal identifications on the blue sublattice are (see Figure \ref{fig:N2parity}(a))
\ie\label{N2ee-id}
(x',y') \sim (x'+\tfrac{L_x}{2},y'-\tfrac{L_x}{2}) \sim (x'+\tfrac{L_y}{2},y'+\tfrac{L_y}{2})~.
\fe
Let $(\tilde L_x,\tilde L_y)$ be the integer solution of the equation $\tilde L_x L_x + \tilde L_y L_y = \gcd(L_x,L_y)$. Then, in the notation of \eqref{minimal-reduced-id}, we have
\ie\label{3danisolap-N2ee-MK}
ML^\text{eff}_{x'} = \lcm(L_x,L_y)~,\qquad KL^\text{eff}_{x'} = \frac12(\tilde L_x L_x - \tilde L_y L_y)~,\qquad L^\text{eff}_{y'} = \frac12\gcd(L_x,L_y)~,
\fe
and hence,
\ie\label{3danisolap-N2ee-gsd}
\text{GSD} = \left[\gcd(2,M) \cdot 2^{L^\text{eff}_{x'} + L^\text{eff}_{y'}-1}\right]^4~.
\fe
Here, the power is $4$ rather than $2$ because there are two copies of the 3+1d anisotropic $\mathbb Z_2$ lineon model, one on each sublattice.

\item \underline{$L_x$ odd, $L_y$ even}: In this case, the identifications in \eqref{3dZ2anisolap-id} couple the two sublattices so that effectively there is only one sublattice, say the blue sublattice in Figure \ref{fig:N2tilt}. Hence, there is only one copy of the the 3+1d anisotropic $\mathbb Z_2$ lineon model. The minimal identifications on the blue sublattice are (see Figure \ref{fig:N2parity}(b))
\ie\label{N2oe-id}
(x,y) \sim (x+2L_x, y) \sim (x,y+L_y)~,
\fe
which can be written as
\ie
(x',y') \sim (x'+L_x,y'-L_x) \sim (x'+\tfrac{L_y}{2},y'+\tfrac{L_y}{2})~,
\fe
in the new coordinates $(x',y')$. Let $(\tilde L_x,\tilde L_y)$ be the integer solution of the equation $\tilde L_x L_x + \tilde L_y (\frac{L_y}2) = \gcd(L_x,\frac{L_y}2)$. Then, in the notation of \eqref{minimal-reduced-id}, we have
\ie\label{3danisolap-N2oe-MK}
ML^\text{eff}_{x'} = 2 \lcm(L_x,\tfrac{L_y}2)~,\qquad KL^\text{eff}_{x'} = \tilde L_x L_x - \frac12\tilde L_y L_y~,\qquad L^\text{eff}_{y'} = \gcd(L_x,\frac{L_y}2)~,
\fe
and hence,
\ie\label{3danisolap-N2oe-gsd}
\text{GSD} = \left[\gcd(2,M) \cdot 2^{L^\text{eff}_{x'} + L^\text{eff}_{y'}-1}\right]^2~.
\fe

\item \underline{$L_x,L_y$ odd}: In this case, once again, the identifications in \eqref{3dZ2anisolap-id} couple the two sublattices so that effectively there is only one sublattice, say the blue sublattice in Figure \ref{fig:N2tilt}. Hence, there is only one copy of the the 3+1d anisotropic $\mathbb Z_2$ lineon model. The minimal identifications on the blue sublattice are (see Figure \ref{fig:N2parity}(c))
\ie\label{N2oo-id}
(x,y) \sim (x+2L_x, y) \sim (x+L_x,y+L_y)~,
\fe
which can be written as
\ie
(x',y') \sim (x'+L_x,y'-L_x) \sim (x'+\tfrac{L_x+L_y}2,y'-\tfrac{L_x-L_y}2)~,
\fe
in the new coordinates $(x',y')$. Let $(\tilde L_x,\tilde L_y)$ be the integer solution of the equation $\tilde L_x L_x + \tilde L_y (\frac{L_x-L_y}2) = \gcd(L_x,\frac{L_x-L_y}2)$. Then, in the notation of \eqref{minimal-reduced-id}, we have
\ie\label{3danisolap-N2oo-MK}
ML^\text{eff}_{x'} = \frac{L_x L_y}{\gcd(L_x,\frac{L_x-L_y}2)}~,\qquad KL^\text{eff}_{x'} = \tilde L_x L_x + \tilde L_y(\tfrac{L_x + L_y}2)~,\qquad L^\text{eff}_{y'} = \gcd(L_x,\tfrac{L_x-L_y}2)~,
\fe
and hence,
\ie\label{3danisolap-N2oo-gsd}
\text{GSD} = \left[\gcd(2,M) \cdot 2^{L^\text{eff}_{x'} + L^\text{eff}_{y'}-1}\right]^2~.
\fe
\end{itemize}
When $L_x = L_y = L$, we have
\ie\label{MKforL}
L^\text{eff}_{x'} = L^\text{eff}_{y'} = \frac{L}{M}~, \qquad K = 1~,\qquad M = \begin{cases}
2~,&L \text{ even}~,
\\
1~,&L \text{ odd}~.
\end{cases}
\fe
Then, the above expressions for the ground state degeneracy simplify to
\ie\label{3dZ2-GSD}
\text{GSD} = \begin{cases}
2^{4L}~,\qquad & L \text{ even}~,
\\
2^{4L-2}~,\qquad & L \text{ odd}~.
\end{cases}
\fe

\subsection{Mobility restrictions}

Let us now discuss the mobility of the $z$-lineons in the $xy$-plane in the 3+1d anisotropic $\mathbb Z_2$ Laplacian model. First, note that in the 2+1d $\mathbb Z_2$ Ising plaquette model on a 2d spatial torus given by the identifications \eqref{minimal-reduced-id}, the defect describing a single particle of unit charge can ``hop'' from $(x',y')$ to $(x'+\gcd(2,M)L^\text{eff}_{x'},y')$ \cite{Gorantla:2022eem}. This motion is nontrivial if and only if $\gcd(2,M) = 1$ and $M>1$. It follows that the $z$-lineons of the 3+1d anisotropic $\mathbb Z_2$ lineon model, in addition to moving along the $z$ direction, can ``hop'' in the $xy$-plane in the same way.

In particular, consider $L_x = L_y = L$. It follows from \eqref{MKforL} that the two conditions  $\gcd(2,M) = 1$ and $M > 1$ cannot be simultaneously satisfied for any $L$. Therefore, the $z$-lineons cannot hop in the $xy$-plane, but can only move in the $z$-direction (and hence the name lineon).

A dipole of $z$-lineons at $(x,y)$ and $(x+s,y\pm s)$, where $s\in\mathbb Z$, can move in the $(1,\mp1)$ direction. This follows from the motion of a dipole of fractons in the 2+1d $\mathbb Z_2$ Ising plaquette model in the direction orthogonal to their separation.

\section{3+1d anisotropic $\mathbb Z_p$ Laplacian model}\label{app:3dZpanisolap}
In this appendix, we use techniques from commutative algebra to compute the ground state degeneracy and analyze the restricted mobility of the 3+1d anisotropic $\mathbb Z_p$ Laplacian model of Section \ref{sec:3dZNanisolap-Np} when $p>2$ is prime. Such techniques were used to analyze translationally invariant Pauli stabilizer codes \cite{Haah_2013}. All the mathematical facts used here can be found in standard textbooks on the subject, such as \cite{dummit2004,eisenbud2013}.

\subsection{Ground state degeneracy}\label{app:3dZpanisolap-gsd}
Recall the relation \eqref{gsd-equivcl} between the GSD of the anisotropic $\mathbb Z_N$ Laplacian model and the number of equivalence classes of $\mathbb Z_N$-valued functions under the equivalence relation ``$\sim$'' of \eqref{graph-equiv}. Here, we set $N=p>2$, where $p$ is prime,\footnote{Actually, all of the following discussion up to \eqref{ZNlap-GSD} works even for $p=2$. The discussion after that does not work for $p=2$ for reasons we will explain later.} and $\Gamma = C_{L_x} \times C_{L_y}$, the 2d torus graph. We first find an exact expression for the $\log_p\text{GSD}$ in terms of commutative-algebraic quantities using the relation \eqref{gsd-equivcl}, then explain how to compute this expression in general using a \emph{Gr\"obner basis}, and finally compute it explicitly for some special values of $L_x,L_y$. In particular, we show that there are infinitely many $L_x,L_y$ for which $\log_p\text{GSD}$ is $O(L_x,L_y)$, and also infinitely many $L_x,L_y$ for which $\log_p\text{GSD}$ is finite.

\subsubsection{Exact expression for $\log_p\mathrm{GSD}$}

Since there are $L_xL_y$ points in the $\Gamma = C_{L_x} \times C_{L_y}$ torus and since we are interested in $\mathbb Z_p$-valued functions on that space, it is clear that there are $p^{L_xL_y}$ such functions.  As in  \eqref{graph-equiv}, they fall into equivalence classes $g(x,y) \sim \tilde g(x,y)$ when $g(x,y) - \tilde g(x,y) = \Delta_L f(x,y) \mod p$. We would like to find the number of such equivalence classes.

As a first step, we give a more abstract description of these $p^{L_xL_y}$ functions. Let $\mathcal R = \mathbb Z_p[X,Y]$ be the ring of polynomials with coefficients in $\mathbb Z_p$, and $\mathfrak j = (\mathsf Q_x,\mathsf Q_y)$ be the \emph{ideal} of $\mathcal R$ generated by the polynomials $\mathsf Q_x(X,Y) = X^{L_x} - 1$ and $\mathsf Q_y(X,Y) = Y^{L_y} - 1$. Given two polynomials $\mathsf F,\mathsf G\in\mathcal R$, we write\footnote{We do not write the ``mod $p$'' explicitly because we are working in $\mathbb Z_p[X,Y]$.}
\ie
\mathsf F(X,Y) = \mathsf G(X,Y) \mod \mathfrak j~,
\fe
if and only if $\mathsf F(X,Y) - \mathsf G(X,Y)$ is a polynomial in $\mathfrak j$. The set of equivalence classes modulo $\mathfrak j$ is the quotient ring $\mathcal R/\mathfrak j$.

Any equivalence class of $\mathcal R/\mathfrak j$ is represented by a unique polynomial that is a $\mathbb Z_p$-linear combination of the monomials $X^aY^b$ with $0\le a<L_x$ and $0\le b<L_y$. (Here, we used the equivalence relations to remove higher powers of $X$ or $Y$.  This is a special case of a more general procedure, called complete reduction, which we will describe below.)  Therefore, the number of equivalence classes is $p^{L_x L_y}$. In fact, since $\mathbb Z_p$ is a field, $\mathcal R/\mathfrak j$ can be thought of as a vector space over $\mathbb Z_p$. The above monomials form a basis of this vector space, so
\ie\label{LxLydim}
\dim_{\mathbb Z_p} \mathcal R/\mathfrak j = L_x L_y~.
\fe
Here, ``$\dim_{\mathbb Z_p}$'' denotes the dimension of a vector space over $\mathbb Z_p$.

It is convenient to represent a $\mathbb Z_p$-valued function $f(x,y)$ on the 2d torus graph $\Gamma = C_{L_x} \times C_{L_y}$ as a polynomial representing an equivalence class of the quotient ring $\mathcal R/\mathfrak j$ as follows:
\ie\label{poly-rep-func}
\hat f(X,Y) = \sum_{x=0}^{L_x-1} \sum_{y=0}^{L_y-1} f(x,y) X^{L_x - x - 1} Y^{L_y - y - 1} \mod \mathfrak j~.
\fe
$\hat f(X,Y)$ can be thought of as a lattice Fourier transform of $f(x,y)$ with $X=e^{ik_x}$ and $Y=e^{ik_y}$, which depend on the momenta $k_x$ and $k_y$.

Observe that, for any integer $0\le k<L_x$, we have
\ie\label{k translations}
X^k\hat f(X,Y) &= \bigg[\sum_{x=0}^{L_x-k-1} \sum_{y=0}^{L_y-1} f(x+k,y) X^{L_x - x-1} Y^{L_y - y - 1}
\\
&\qquad + \sum_{x=L_x-k}^{L_x-1}\sum_{y=0}^{L_y-1} f(x-L_x+k,y) X^{L_x - x-1}Y^{L_y - y - 1} \bigg] \mod \mathfrak j~,
\fe
so $X$ can be interpreted as the generator of translations in the $x$ direction.\footnote{Our functions $f(x,y)$ are defined on $L_xL_y$ points.  They can be extended to periodic functions on $\mathbb Z^2$.  Then, it is straightforward to apply $k$ translations.  The expression \eqref{k translations} corresponds to applying such $k$ translation and then expressing the result in terms of the function $f(x,y)$ in the fundamental domain $0\le x \le L_x-1$, $0\le y \le L_y-1$.  Equivalently, we can use the periodicity of $f$ to write \eqref{k translations} as $X^k\hat f(X,Y) = \sum_{x=0}^{L_x-1} \sum_{y=0}^{L_y-1} f(x+k,y) X^{L_x - x-1} Y^{L_y - y - 1} \mod \mathfrak j$.}  The fact that translating in $x$ by $L_x$ takes the graph $C_{L_x}$ back to itself is related to the trivial equation
\ie
X^{L_x} \hat f(X,Y) = \hat f(X,Y) \mod \mathfrak j~.
\fe
Also, the difference operator $\Delta_x$ is associated with the polynomial $X-1$. For convenience, we define the \emph{displaced} discrete Laplacian operator, denoted by $\tilde \Delta_L$, as
\ie
\tilde \Delta_L f(x,y) = (\Delta_x^2+\Delta_y^2) f(x+1,y+1)~.
\fe
(Here, we extended $f(x,y)$ to a periodic function on $\mathbb Z^2$.) It is associated with the polynomial
\ie\label{disclap-poly}
\tilde{\mathsf P}(X,Y) = Y(X-1)^2 + X(Y-1)^2~.
\fe
In general, any \emph{local} difference operator in the $xy$-plane, after an appropriate displacement, is associated with a polynomial $\mathsf S\in\mathcal R$ satisfying $\mathsf S(1,1) = 0$.

Let $\mathfrak i = (\tilde{\mathsf P})$ be the ideal of $\mathcal R$ generated by the polynomial $\tilde{\mathsf P}(X,Y)$. Then, $\mathfrak i/(\mathfrak i \cap \mathfrak j) \cong (\mathfrak i + \mathfrak j)/\mathfrak j$ is an ideal of the quotient ring $\mathcal R/\mathfrak j$.\footnote{Here, $\mathfrak i + \mathfrak j = (\tilde{\mathsf P},\mathsf Q_x,\mathsf Q_y)$ is the ideal of $\mathcal R$ generated by the three polynomials $\tilde{\mathsf P}(X,Y)$, $\mathsf Q_x(X,Y)$, and $\mathsf Q_y(X,Y)$. It is known as the \emph{sum of ideals} $\mathfrak i$ and $\mathfrak j$.} In fact, it is the subspace of the vector space $\mathcal R/\mathfrak j$ that corresponds to the $\mathbb Z_p$-valued functions of the form $\tilde \Delta_L f(x,y)$.

It is then clear that the quotient
\ie\label{quotient-equiv}
\frac{\mathcal R/\mathfrak j}{(\mathfrak i + \mathfrak j)/\mathfrak j} \cong \mathcal R/(\mathfrak i + \mathfrak j)~,
\fe
corresponds to the set of equivalence classes of $\mathbb Z_p$-valued functions under the equivalence relation ``$\sim$'' of \eqref{graph-equiv} on the 2d torus graph $C_{L_x}\times C_{L_y}$. It follows that
\ie\label{app:gsd-exact}
\log_p \text{GSD} = 2 \dim_{\mathbb Z_p} \mathcal R/(\mathfrak i + \mathfrak j)~,
\fe
or more explicitly,
\ie
\log_p \text{GSD} = 2 \dim_{\mathbb Z_p} \frac{\mathbb Z_p[X,Y]}{(Y(X-1)^2 + X(Y-1)^2,X^{L_x}-1,Y^{L_y}-1)}~.
\fe
Below, we give a general procedure to compute this quantity.

\subsubsection{Computing $\log_p \mathrm{GSD}$ using Gr\"obner basis}
One way to compute the (vector space) dimension of $\mathcal R/(\mathfrak i + \mathfrak j)$ is by first computing the \emph{Gr\"obner basis} of the ideal $\mathfrak i + \mathfrak j$. Before defining a Gr\"obner basis, we need an ordering on all the monomials. For our purposes, it is sufficient to define a \emph{lexicographic monomial ordering}: $X^mY^n \succ X^k Y^l$ if and only if $m>k$, or $m=k$ and $n>l$. Then, for any polynomial $\mathsf F(X,Y)$, we define its \emph{leading term} as the term that has the largest monomial among all the terms.

Now, we say a polynomial $\mathsf F(X,Y)$ is \emph{reducible} with respect to a set of polynomials $\mathscr G = \{\mathsf G_1,\ldots,\mathsf G_n\}$ if some term of $\mathsf F(X,Y)$ is a multiple of the leading term of one of the $\mathsf G_i(X,Y)$'s. We say it is \emph{irreducible} otherwise.

Given an ideal $\mathfrak I$ of $\mathcal R$ and a polynomial $\mathsf F\in\mathcal R$, one can ask what the equivalence class of $\mathsf F(X,Y)$ in $\mathcal R/\mathfrak I$ is. One way to answer this question is to \emph{reduce} $\mathsf F(X,Y)$ with respect to a generating set $\mathscr B$ of $\mathfrak I$ repeatedly until we are left with a polynomial $\mathsf H(X,Y)$ that is irreducible with respect to $\mathscr B$. (Such a procedure is known as a \emph{complete reduction} of $\mathsf F(X,Y)$ with respect to $\mathscr B$.) One might hope that $\mathsf H(X,Y)$ uniquely specifies the equivalence class of $\mathsf F(X,Y)$. However, for a generic $\mathscr B$, the $\mathsf H(X,Y)$ so obtained depends on the choices made in the repeated reductions, and hence, may not be unique.

A Gr\"obner basis $\mathscr G = \{\mathsf G_1,\ldots,\mathsf G_n\}$ is a special generating set of $\mathfrak I$ such that $\mathsf F(X,Y)$ can be written as
\ie\label{grobner-multivardiv}
\mathsf F(X,Y) = \mathsf H(X,Y) + \sum_{i=1}^n \mathsf H_i(X,Y) \mathsf G_i(X,Y)~,
\fe
where $\mathsf H(X,Y)$ is \emph{uniquely} determined by the requirement that it is irreducible with respect to $\mathscr G$.\footnote{Note that $\mathsf H_i(X,Y)$'s are not uniquely determined by this procedure. Indeed, shifting $\mathsf H_1(X,Y)$ by $\mathsf G_2(X,Y)$ and $\mathsf H_2(X,Y)$ by $-\mathsf G_1(X,Y)$ gives another complete reduction of $\mathsf F(X,Y)$ with respect to $\mathscr G$.} In this case, we write
\ie
\mathsf F(X,Y) = \mathsf H(X,Y) \mod \mathscr G~.
\fe
It follows that there is a one-one correspondence between $\mathcal R/\mathfrak I$ and the set of all polynomials that are irreducible with respect to $\mathscr G$. Indeed, the set of all monomials that are irreducible with respect to $\mathscr G$ forms a basis of the vector space $\mathcal R/\mathfrak I$. From this we conclude that $\dim_{\mathbb Z_p} \mathcal R/\mathfrak I$ equals the number of monomials that are irreducible with respect to $\mathscr G$.

While there is an algorithm, known as \emph{Buchberger's algorithm}, to compute a Gr\"obner basis of an ideal given its generators, it is not always easy to compute it analytically.\footnote{For the ideal $\mathfrak j = (\mathsf Q_x,\mathsf Q_y)$, the generating set $\{\mathsf Q_x,\mathsf Q_y\}$ is already a Gr\"obner basis, and moreover, the irreducible monomials are $X^x Y^y$ for $0\le x< L_x$ and $0\le y< L_y$, which form a basis of $\mathcal R/\mathfrak j$. (This fact was used in the analysis leading to \eqref{LxLydim}.) However, for the ideal $\mathfrak i + \mathfrak j = (\tilde{\mathsf P},\mathsf Q_x,\mathsf Q_y)$, the generating set $\{\tilde{\mathsf P},\mathsf Q_x,\mathsf Q_y\}$ is not always a Gr\"obner basis.}
Nonetheless, for fixed values of $p,L_x,L_y$, the Gr\"obner basis, and therefore the GSD, can be readily computed with the help of computer programs.
Let us do an explicit calculation for the GSD when $p=3$ and $L_x=L_y=4$ as an example.
Using the $\mathsf{GroebnerBasis}$ command with $\mathsf{Modulus}\rightarrow 3$ in $\mathsf{Mathematica}$,
we can compute the Gr\"obner basis for the ideal $(\tilde{\mathsf P},\mathsf Q_x,\mathsf Q_y)= \left(Y(X-1)^2+X(Y-1)^2 ,X^{4}-1,Y^4-1\right)$ in this case. We find $\mathscr G = \{ Y^4-1 ,XY+X-Y^3-Y^2,X^2 +Y^3-Y^2+Y+1\}$ under the lexicographic ordering  $X\succ Y$. The leading terms of $\mathscr G$ are $\{Y^4 , XY,X^2\}$, and  the 5 irreducible monomials with respect to $\mathscr G$ are $1,Y,Y^2,Y^3,X$.
We conclude that $ \dim_{\mathbb Z_3} \mathcal R/(\mathfrak i + \mathfrak j) = 5$, and hence $\log_3 \text{GSD} = 10$. In fact, the plot of $\log_N \text{GSD}$ as a function of $L_x = L_y = L$ in Figure \ref{fig:intro-gsd} was obtained exactly in this way.

Fortunately, for the problem at hand, it is possible to simplify the expression \eqref{app:gsd-exact} for the GSD further analytically. To prepare for the simplified expression, we first define a few things:

A field $F$ is \emph{algebraically closed} if any polynomial in $F[X]$ can be factorized completely into linear factors in $F[X]$. For example, $\mathbb R$ is not algebraically closed because the polynomial $X^2+1$, which is in $\mathbb R[X]$, cannot be factorized into linear factors in $\mathbb R[X]$. In contrast, it is well-known that $\mathbb C$ is algebraically closed. Every field, by definition, contains the multiplicative identity $1$.

The \emph{characteristic} of a field is the smallest positive integer $n$ such that $1 + \cdots + 1 \text{($n$ times)} = 0$. It is clear that $\mathbb Z_p$ is a field of characteristic $p$. By convention, the characteristics of $\mathbb Q$, $\mathbb R$, and $\mathbb C$ are all defined to be $0$. It is known that the characteristic of any field is either $0$ or a prime number. As we see, there are several fields with the same characteristic $p$.

For \emph{finite fields}, the field is uniquely specified by the number of its elements and the characteristic.  In particular, for any integer $k\ge1$ and prime $p$, $\mathbb F_{p^k}$ denotes the unique (up to field isomorphisms) finite field of order $p^k$ and characteristic $p$. Furthermore, $\mathbb F_{p^\infty}$ denotes the unique algebraically closed field of characteristic $p$.\footnote{More explicitly, $\mathbb F_{p^\infty} = \bigcup_{n=1}^\infty \mathbb F_{p^{n!}}$. Here we used the fact that $\mathbb F_{p^k}$ is a subfield of $\mathbb F_{p^m}$ if and only if $k$ divides $m$.}

Let $\mathcal S = \mathbb F_{p^\infty}[X,Y]$ be the ring of polynomials in $X,Y$ with coefficients in $\mathbb F_{p^\infty}$. We use the same symbols $\mathfrak i$ and $\mathfrak j$ for the ideals of $\mathcal S$ generated by the polynomials $\tilde{\mathsf P}(X,Y)$ and $\mathsf Q_x(X,Y),\mathsf Q_y(X,Y)$ respectively. The \emph{algebraic set} of $\mathfrak i + \mathfrak j$, denoted as $V(\mathfrak i + \mathfrak j)$, is the set of distinct solutions $(X_0,Y_0)\in \mathbb F_{p^\infty}^2$ of the system of polynomial equations
\ie
X^{L_x} - 1 = Y^{L_y} - 1 = \tilde{\mathsf P}(X,Y) = 0~.
\fe

We will find it convenient to parameterize $L_x$ and $L_y$ in terms of the order of our group $\mathbb Z_p$ as
\ie\label{Lprimed}
&L_x = p^{k_x} L_x'~,\qquad L_y = p^{k_y} L_y'~,\qquad\gcd(p,L_x')=\gcd(p,L_y')=1~.
\fe
Now, for each solution $(X_0,Y_0)\in V(\mathfrak i + \mathfrak j)$, we define the polynomials
\ie
\tilde{\mathsf Q}_{x,X_0}(X,Y) = (X-X_0)^{p^{k_x}}~,\qquad \tilde{\mathsf Q}_{y,Y_0}(X,Y) = (Y-Y_0)^{p^{k_y}}~,
\fe
and the ideal $\mathfrak i'_{X_0,Y_0} = (\tilde{\mathsf P},\tilde{\mathsf Q}_{x,X_0},\tilde{\mathsf Q}_{y,Y_0})$ of $\mathcal S$.

With these preparations, the simplified expression for the GSD \eqref{app:gsd-exact} is
\ie\label{ZNlap-GSD-exact}
\log_p\text{GSD} = 2 \sum_{(X_0,Y_0)\in V(\mathfrak i + \mathfrak j)}  \dim_{\mathbb F_{p^\infty}}\mathcal S/\mathfrak i'_{X_0,Y_0}~.
\fe
It is obtained using techniques from commutative algebra (see e.g., \cite{dummit2004,eisenbud2013}), that were used in \cite{Haah_2013}. For readers who are familiar with such techniques, a derivation of \eqref{ZNlap-GSD-exact} is given below. Others, who are willing to accept it, can skip directly to Appendix \ref{app:specialGSD}, where we compute the GSD for some special values of $L_x,L_y$ explicitly.

\begin{center}
\emph{Derivation of \eqref{ZNlap-GSD-exact}}
\end{center}
Since $\mathcal R/(\mathfrak i + \mathfrak j)$ is a finite-dimensional vector space over $\mathbb Z_p$, it is an \emph{Artinian} ring,\footnote{A ring $\mathcal R$ is Artinian if it satisfies the \emph{descending chain condition}, i.e., if $\mathfrak I_1 \supseteq \mathfrak I_2 \supseteq \mathfrak I_3 \supseteq \cdots$ is a descending chain of ideals, then there is a $k\ge1$ such that $\mathfrak I_k = \mathfrak I_{k+1} = \mathfrak I_{k+2} = \cdots$. For example, the ring of integers $\mathbb Z$ is not Artinian because $(2) \supset (4) \supset (8) \supset \cdots$, where $(n)$ denotes the ideal generated by the integer $n$. On the other hand, for any integer $n$, the ring of integers modulo $n$, $\mathbb Z/(n) = \mathbb Z_n$, is Artinian. Moreover, any ring that is also a finite dimensional vector space over a field is always Artinian, which is exactly the case here.} and hence, it has finitely many \emph{maximal ideals}.\footnote{A proper ideal is an ideal that is not the ring itself. For example, in $\mathbb Z$, the ideal $(4)$ is proper because it does not contain $1$. (The only ideal containing $1$ is the entire ring itself.) A maximal ideal is a proper ideal that is not contained in any other proper ideal except itself. For example, in $\mathbb Z$, $(4)$ is not a maximal ideal because it is contained in the proper ideal $(2)$. The latter is maximal; in fact, the ideal $(n)$ is maximal if and only if $n$ is prime.} Moreover, for an Artinian ring, it is known that
\ie\label{max-id-sum}
\mathcal R/(\mathfrak i + \mathfrak j) \cong \bigoplus_{\mathfrak m} [\mathcal R/(\mathfrak i + \mathfrak j)]_{\mathfrak m}~,
\fe
where the sum is over all maximal ideals of $\mathcal R/(\mathfrak i + \mathfrak j)$, and $[\mathcal R/(\mathfrak i + \mathfrak j)]_{\mathfrak m}$ denotes the \emph{localization} of $\mathcal R/(\mathfrak i + \mathfrak j)$ at $\mathfrak m$.\footnote{Intuitively, given a multiplicatively closed subset $S$ of a ring $\mathcal R$, the localization of $\mathcal R$ with respect to $S$, denoted as $S^{-1} \mathcal R$, means ``formally adding multiplicative inverses'' for all the elements of $S$. For example, in $\mathbb Z$, the subset of nonzero integers is multiplicatively closed, and localizing with respect to this set gives the rationals $\mathbb Q$. In any ring $\mathcal R$, given a maximal ideal $\mathfrak m$, the set $\mathcal R\setminus \mathfrak m$ is always multiplicatively closed. So we define the localization of $\mathcal R$ at $\mathfrak m$, denoted by $\mathcal R_{\mathfrak m}$, as the localization of $\mathcal R$ with respect to $\mathcal R\setminus \mathfrak m$.} It follows that
\ie
\log_p\text{GSD} = 2\sum_{\mathfrak m} \dim_{\mathbb Z_p} [\mathcal R/(\mathfrak i + \mathfrak j)]_{\mathfrak m}~.
\fe
So, we can compute dimension for each term in the sum and then add them up. However, the maximal ideals of $\mathcal R/(\mathfrak i + \mathfrak j)$ are a bit complicated to work with. Instead, we proceed as follows.

We can replace $\mathbb Z_p$ by $\mathbb F_{p^\infty}$ in \eqref{app:gsd-exact} and get the same answer for $\log_p\text{GSD}$. More concretely, we have\footnote{Given a field $F$, one can talk about extending it to a larger field $F'$ such that $F$ is a subfield of $F'$. For example, $\mathbb R$ is a subfield of $\mathbb C$, or equivalently, $\mathbb C$ is an extension of $\mathbb R$. Now, let $V$ be a vector space over $F$. One can ``extend the base field'' from $F$ to $F'$ by tensoring $V$ with $F'$, denoted as $V\otimes_F F'$. Then, $\dim_F V = \dim_{F'} (V \otimes_F F')$. Since $\mathbb F_{p^\infty}$ is an extension of $\mathbb Z_p$, the result in \eqref{extension} follows.}
\ie\label{extension}
\dim_{\mathbb Z_p} \mathcal R/(\mathfrak i + \mathfrak j) = \dim_{\mathbb F_{p^\infty}} \mathcal S/(\mathfrak i + \mathfrak j)~,
\fe
where, on the right hand side, $\mathfrak i$ and $\mathfrak j$ are ideals of $\mathcal S$ generated by the same polynomials as before. Since $\mathcal S/(\mathfrak i + \mathfrak j)$ is also Artinian, we have
\ie
\log_p\text{GSD} = 2\sum_{\mathfrak m} \dim_{\mathbb F_{p^\infty}} [\mathcal S/(\mathfrak i + \mathfrak j)]_{\mathfrak m}~,
\fe
where the sum is over all maximal ideals of $\mathcal S/(\mathfrak i + \mathfrak j)$. We now characterize the maximal ideals of $\mathcal S/(\mathfrak i + \mathfrak j)$.

By the \emph{correspondence theorem for quotient rings}, the maximal ideals of $\mathcal S/(\mathfrak i + \mathfrak j)$ are in one-one correspondence with the maximal ideals of $\mathcal S$ that contain the ideal $\mathfrak i + \mathfrak j$. Moreover, since $\mathbb F_{p^\infty}$ is algebraically closed, by \emph{Hilbert's Nullstellensatz}, the maximal ideals of $\mathcal S$ are in one-one correspondence with ideals of the form $(X-X_0,Y-Y_0)$, where $(X_0,Y_0)\in\mathbb F_{p^\infty}^2$. Now, the ideal $(X-X_0,Y-Y_0)$ contains $\mathfrak i + \mathfrak j$ if and only if $(X_0,Y_0)$ is a root of all the polynomials in $\mathfrak i + \mathfrak j$. It follows that the maximal ideals of $\mathcal S/(\mathfrak i + \mathfrak j)$ are in one-one correspondence with ideals of $\mathcal S$ of the form $(X-X_0,Y-Y_0)$, where $(X_0,Y_0)$ is a solution of the system of polynomial equations
\ie\label{poly-sysold}
X^{L_x}-1 = Y^{L_y}-1=\tilde{\mathsf P}(X,Y)=0~,
\fe
i.e., $(X_0,Y_0) \in V(\mathfrak i + \mathfrak j)$, the algebraic set of $\mathfrak i + \mathfrak j$. Then,
\ie
\log_p\text{GSD} = 2 \sum_{(X_0,Y_0)\in V(\mathfrak i + \mathfrak j)} \dim_{\mathbb F_{p^\infty}} [\mathcal S/(\mathfrak i + \mathfrak j)]_{(X-X_0,Y-Y_0)}~.
\fe

We can simplify it further. Let $L_i = p^{k_i} L_i'$ with $\gcd(p,L_i')=1$. Recall that since $\mathbb F_{p^\infty}$ is algebraically, any polynomial in $\mathbb F_{p^\infty}[X]$ can be factorized completely into linear factors in $\mathbb F_{p^\infty}[X]$. Using this fact, we have
\ie\label{lin-factor}
X^{L_x} -1 = (X^{L_x'}-1)^{p^{k_x}} = \prod_{\xi \in \mathbb F_{p^\infty}~:~\xi^{L_x'} = 1} (X-\xi)^{p^{k_x}}~,
\fe
That is, there are $L_x'$ distinct $\xi$'s in $\mathbb F_{p^\infty}$ that satisfy $\xi^{L_x}=1$, each with multiplicity $p^{k_x}$.

Let $(X_0,Y_0) \in V(\mathfrak i + \mathfrak j)$. It is clear that $X_0$ is one of the $\xi$'s in \eqref{lin-factor}. Consider the localization $\mathcal S_{(X-X_0,Y-Y_0)}$. Recall that every element outside the ideal $(X-X_0,Y-Y_0)$ becomes a \emph{unit} in the localization $\mathcal S_{(X-X_0,Y-Y_0)}$.\footnote{A unit is an element of the ring that has a multiplicative inverse. In any ring, the multiplicative identity $1$ is its own inverse, so it is always a unit. In particular, in $\mathbb Z$, $\pm1$ are the only units, whereas in $\mathbb Q$, any nonzero rational is a unit.} In particular, the polynomial $\mathsf Q_x(X,Y) = X^{L_x}-1$ generates the same ideal in $\mathcal S_{(X-X_0,Y-Y_0)}$ as the polynomial $\tilde{\mathsf Q}_{x,X_0}(X,Y) = (X-X_0)^{p^{k_x}}$ because the other linear factors in \eqref{lin-factor} associated with $\xi\ne X_0$ all have inverses.\footnote{More abstractly, if $r$ is one of the generators of an ideal $\mathfrak I$ of $\mathcal R$, then we can replace it by $ur$, where $u$ is a unit.} Similarly, $\mathsf Q_y(X,Y) = Y^{L_y}-1$ generates the same ideal as $\tilde{\mathsf Q}_{y,Y_0}(X,Y) = (Y-Y_0)^{p^{k_y}}$. It follows that
\ie\label{same-ideal-loc}
(\mathfrak i + \mathfrak j)_{(X-X_0,Y-Y_0)} = (\tilde{\mathsf P},\tilde{\mathsf Q}_{x,X_0},\tilde{\mathsf Q}_{y,Y_0})_{(X-X_0,Y-Y_0)}~,
\fe
in the localization $\mathcal S_{(X-X_0,Y-Y_0)}$. Here, we use the notation $\mathfrak I_{\mathfrak m}$ to denote the ideal in $\mathcal S_{\mathfrak m}$ generated by the image of the ideal $\mathfrak I \subseteq \mathcal S$ under the \emph{localization map} $\mathcal S \rightarrow \mathcal S_{\mathfrak m}$, where $\mathfrak m$ is a maximal ideal of $\mathcal S$.

Defining $\mathfrak i'_{X_0,Y_0} = (\tilde{\mathsf P},\tilde{\mathsf Q}_{x,X_0},\tilde{\mathsf Q}_{y,Y_0})$, we then have
\ie
[\mathcal S/(\mathfrak i + \mathfrak j)]_{(X-X_0,Y-Y_0)} &\cong \mathcal S_{(X-X_0,Y-Y_0)}/(\mathfrak i + \mathfrak j)_{(X-X_0,Y-Y_0)}
\\
&\cong \mathcal S_{(X-X_0,Y-Y_0)}/(\mathfrak i'_{X_0,Y_0})_{(X-X_0,Y-Y_0)}
\\
&\cong (\mathcal S/\mathfrak i'_{X_0,Y_0})_{(X-X_0,Y-Y_0)}~,
\fe
where in the first and third lines, we used the slogan ``localization commutes with quotienting'', and the second line follows from \eqref{same-ideal-loc}. Now, the quotient $\mathcal S/\mathfrak i'_{X_0,Y_0}$ is also Artinian, and by Hilbert's Nullstellensatz, its only maximal ideal is $(X-X_0,Y-Y_0)$. Hence, by a result similar to the one in \eqref{max-id-sum}, we have
\ie
(\mathcal S/\mathfrak i'_{X_0,Y_0})_{(X-X_0,Y-Y_0)} \cong \mathcal S/\mathfrak i'_{X_0,Y_0}~,
\fe
and therefore,
\ie\label{ZNlap-GSD-exact2}
\log_p\text{GSD} = 2 \sum_{(X_0,Y_0)\in V(\mathfrak i + \mathfrak j)}  \dim_{\mathbb F_{p^\infty}}\mathcal S/\mathfrak i'_{X_0,Y_0}~.
\fe

\subsubsection{Computation of $\log_p\mathrm{GSD}$ for special values of $L_x,L_y$}\label{app:specialGSD}
Given the expression \eqref{ZNlap-GSD-exact} for the GSD, all we need to do now is to compute a Gr\"obner basis of $\mathfrak i'_{X_0,Y_0}$ in $\mathcal S$ for each $(X_0,Y_0) \in V(\mathfrak i + \mathfrak j)$. First, we note certain ``symmetries'' in the set $V(\mathfrak i + \mathfrak j)$. Recall that $V(\mathfrak i + \mathfrak j)$ is the set of $(X_0,Y_0)\in\mathbb F_{p^\infty}^2$ that solve the system of polynomial equations
\ie\label{poly-sys}
X^{L_x'}-1 = Y^{L_y'}-1=\tilde{\mathsf P}(X,Y)=0~,
\fe
where we used the facts that $X^{L_x}-1 = (X^{L_x'}-1)^{p^{k_x}}$ and $Y^{L_y}-1 = (Y^{L_y'}-1)^{p^{k_y}}$ in $\mathbb F_{p^\infty}^2[X,Y]$. (Here, we used the parametrization \eqref{Lprimed}.) Given a solution $(X_0,Y_0)$ of \eqref{poly-sys}, we can generate three more solutions using the transformations
\ie\label{transinv}
X_0 \rightarrow X_0^{-1}~,\qquad Y_0 \rightarrow Y_0^{-1}~,
\fe
because the equations \eqref{poly-sys} are invariant under these transformations. (These transformations are well defined because $X_0 \ne 0$ and $Y_0\ne0$.)  Furthermore, if $L_x' = L_y'$, we can generate four more solutions using the exchange
\ie\label{transexch}
X_0 \leftrightarrow Y_0~.
\fe
It is clear that $(1,1)$ is the only solution that is invariant under all these transformations. We will exploit these transformations in our analysis below.

In general, it is hard to compute a Gr\"obner basis of $\mathfrak i'_{X_0,Y_0}$ for arbitrary $(X_0,Y_0)\in\mathbb F_{p^\infty}^2$ except when $(X_0,Y_0) = (1,1)$. So, below, we specialize to those values of $L_x,L_y$ for which $(X_0,Y_0) = (1,1)$ is the only solution of \eqref{poly-sys}. These special values of $L_x,L_y$ contain infinite families of $L_x,L_y$ with interesting behaviors of GSD.

\begin{enumerate}
\item Consider the special case $L_x'=L_y'=1$, where \eqref{poly-sys} becomes
\ie\label{poly-sys1}
X-1 = Y-1=\tilde{\mathsf P}(X,Y)=0~.
\fe
Clearly, $(X_0,Y_0)=(1,1)$ is the only solution of \eqref{poly-sys1}. In this case, changing the variables $X$ and $Y$ to $\tilde X=X-1$ and $\tilde Y=Y-1$, we have $\mathfrak i'_{1,1} = (\tilde X^2\tilde Y+\tilde X\tilde Y^2+\tilde X^2+\tilde Y^2,\tilde X^{p^{k_x}},\tilde Y^{p^{k_y}})$. We can assume that $k_x \ge k_y$ without loss of generality. Then, with a lexicographic monomial order on $\tilde X,\tilde Y$ with $\tilde X\succ \tilde Y$, a Gr\"obner basis of $\mathfrak i'_{1,1}$ is given by the following:
\begin{itemize}
\item When $k_y > 0$,
\ie
&\mathsf G_1(\tilde X,\tilde Y) = \tilde Y^{p^{k_y}}~,
\\
&\mathsf G_2(\tilde X,\tilde Y) = \tilde X \tilde Y^{p^{k_y} - 1} \delta_{k_x,k_y}~,
\\
&\mathsf G_3(\tilde X,\tilde Y) = \tilde X^2 + (\tilde X+1)\tilde Y^2 \sum_{i=0}^{p^{k_y}-3} (-\tilde Y)^i - \tilde X \tilde Y^{p^{k_y} - 1} \delta_{k_x,k_y}~.
\fe
\item When $k_y=0$,
\ie
&\mathsf G_1(\tilde X,\tilde Y) = \tilde Y~,
\\
&\mathsf G_2(\tilde X,\tilde Y) = \tilde X \delta_{k_x,0} + \tilde X^2 (1-\delta_{k_x,0})~.
\fe
\end{itemize}
The set of monomials that are irreducible with respect to this Gr\"obner basis are
\ie
\{ \tilde Y^i : 0\le i < p^{k_y} \} \cup \{ \tilde X \tilde Y^j: 0 \le j < p^{k_y} - \delta_{k_x,k_y} \}~.
\fe
They form a basis of $\mathcal S/\mathfrak i'_{1,1}$, so we have
\ie
\dim_{\mathbb F_{p^\infty}} \mathcal S/\mathfrak i'_{1,1} = 2p^{k_y} - \delta_{k_x,k_y}~.
\fe
By \eqref{ZNlap-GSD-exact}, we conclude that
\ie\label{ZNlap-GSD}
\log_p\text{GSD} = 2\left(2p^{k_y} - \delta_{k_x,k_y}\right)~.
\fe

\item Next, we generalize the previous special case to $L_x' = L_y' = q$ with $q>2$ a prime such that $p$ is a \emph{primitive root} modulo $q$.\footnote{In other words, $p$ is a generator of the multiplicative group of integers modulo $q$, denoted as $\mathbb Z_q^\times$. For any positive integer $n$, the order of the group $\mathbb Z_n^\times$ is known as the \emph{Euler totient function} of $n$, denoted as $\varphi(n)$. It is easy to see that $\varphi(q^m)=q^m-q^{m-1}$ for any odd prime $q$ and any $m\ge1$.\label{ftnt:primroot}} Then, \eqref{poly-sys} becomes
\ie\label{poly-sys2}
X^q-1 = Y^q-1=\tilde{\mathsf P}(X,Y)=0~.
\fe
We will show that $X_0=Y_0=1$ is the only solution of \eqref{poly-sys2}.  We argue by contradiction. We assume that there is a solution with $X_0 \ne 1$. Then, any other solution $(X_0,Y_0)$ of \eqref{poly-sys2} is obtained from a solution of the form $(X_0,X_0^s)$ for some $1\le s \le (q-1)/2$ using the transformations \eqref{transinv} and \eqref{transexch}.\footnote{Since $X_0 \ne 1$, $X_0^q = 1$, and $q$ is prime, $X_0$ is a primitive $q$th root of unity, i.e., powers of $X_0$ generates all the $q$th roots of unity. Since $Y_0^q=1$, it is a $q$th root of unity, and hence, $Y_0=X_0^s$ for some $0\le s< q$. But $s\ne0$ because there is no solution of \eqref{poly-sys2} of the form $(X_0,1)$ for $X_0\ne1$. Moreover, using the transformation \eqref{transinv}, $(X_0,X_0^{q-s})$ is also a solution, so we can restrict $s$ to the range $1\le s\le (q-1)/2$.} Since $X_0^q = 1$ and $X_0\ne 1$, $X_0$ is a root of the \emph{cyclotomic polynomial} $\Phi_q(X) = \sum_{j=0}^{q-1} X^j$. Moreover, since $(X_0,X_0^s)$ satisfies $\tilde{\mathsf P}(X,Y)=0$, $X_0$ is also a root of the polynomial $\tilde{\mathsf P}(X,X^s)$. We can write
\ie
\tilde{\mathsf P}(X,X^s) = X(X^s-1)^2 + X^s(X-1)^2 = X(X-1)^2\hat{\mathsf P}_s(X)~,
\fe
where
\ie\label{P_s-poly}
\hat{\mathsf P}_s(X) = \left(\textstyle\sum_{i=0}^{s-1} X^i\right)^2 + X^{s-1}~.
\fe
$\hat{\mathsf P}_s(X)$ is a nonzero polynomial in $\mathbb Z_p[X]$ because $\hat{\mathsf P}_s(0) = 1 \mod p$ for $s>1$, and $2 \mod p$ for $s=1$.\footnote{This is not true for $p=2$ because $\hat{\mathsf P}_s(X) = 0$ identically for $s=1$, so there is always a solution of \eqref{poly-sys} of the form $(X_0,X_0)$ even for $X_0\ne1$.\label{p=2fail}} Since $X_0$ is a root of $\tilde{\mathsf P}(X,X^s)$ and $X_0\ne 0,1$, it must be a root of $\hat{\mathsf  P}_s(X)$. We now use the fact that $\Phi_q(X)$ is the \emph{minimal polynomial} of $X_0$ in $\mathbb Z_p[X]$ because $p$ is a primitive root modulo $q$ \cite[Section~11.2.B]{cox2012}. This means $\Phi_q(X)$ divides $\hat{\mathsf P}_s(X)$. But this is impossible because
\ie
\deg_X \hat{\mathsf P}_s(X) = 2s-2 \le q-3 < q-1 = \deg_X \Phi_q(X)~,
\fe
So, there is no such $X_0$. In other words, when $p$ is a primitive root modulo $q$, the only solution of \eqref{poly-sys2} is $(1,1)$. Then, by the analysis in point 1 above, $\log_p\text{GSD}$ is again given by \eqref{ZNlap-GSD}.

\item We now generalize as follows. Let $q>2$ be a prime such that $p$ is a primitive root modulo $q^m$ for some $m\ge2$. Set $L_x' = L_y' = q^m$, so that \eqref{poly-sys} becomes
\ie\label{poly-sys3}
X^{q^m}-1 = Y^{q^m}-1=\tilde{\mathsf P}(X,Y)=0~.
\fe
Again, we argue by contradiction that the only solution of \eqref{poly-sys3}  $(X_0,Y_0)\in\mathbb F_{p^\infty}^2$ is $(1,1)$. We assume that $X_0 \ne 1$. Then, any other solution $(X_0,Y_0)$ of \eqref{poly-sys3} is obtained from a solution of the form $(X_0,X_0^s)$ for some $1\le s\le (q^m-1)/2$ using the transformations \eqref{transinv} and \eqref{transexch}. Actually, the range of $s$ can be smaller than this. Let $q^r$ be the order of $X_0$ for some $0\le r\le m$, i.e., $X_0^{q^r} = 1$, but $X_0^{q^{r'}} \ne 1$ for any $r'<r$. Since $r=0$ corresponds to the trivial solution $(1,1)$, we have $r>0$. Then, $1\le s\le (q^r-1)/2$. We consider two cases:
\begin{itemize}
\item \underline{$s \le \varphi(q^r)/2$}: [See footnote \ref{ftnt:primroot} for the definition of $\varphi(q^r)$.] Since the order of $X_0$ is $q^r$, it is a root of the cyclotomic polynomial $\Phi_{q^r}(X) = \sum_{j=0}^{q-1} X^{j q^{r-1}}$. Since $X_0 \ne 0,1$, it is also a root of the polynomial $\hat{\mathsf P}_s(X)$ in \eqref{P_s-poly}. We now use the fact that $\Phi_{q^r}(X)$ is the minimal polynomial of $X_0$ in $\mathbb Z_p[X]$ because $p$ is a primitive root modulo $q^r$ \cite[Section~11.2.B]{cox2012}.\footnote{Here, we used the fact: $p$ is a primitive root modulo $q^m \implies p$ is a primitive root modulo $q^r$ for all $r \le m$. In fact, for $m\ge 2$: $p$ is a primitive root modulo $q^m \implies p$ is a primitive root modulo $q^{m+1}$. Combining these facts: $p$ is a primitive root modulo $q^2 \implies p$ is a primitive root modulo $q^m$ for all $m\ge1$ \cite[Section~2.8]{niven1991}.\label{ftnt:primroot2}} This means $\Phi_{q^r}(X)$ must divide $\hat{\mathsf P}_s(X)$. But this is impossible because
\ie
\deg_X \hat{\mathsf P}_s(X) = 2s-2 \le \varphi(q^r)-2 < \varphi(q^r) = \deg_X \Phi_{q^r}(X)~.
\fe
So, there is no such $X_0$.

\item \underline{$\varphi(q^r)/2 < s \le (q^r-1)/2$}: Let $X_0 = Z_0^2$, where $Z_0$ also has order $q^r$. (Such a $Z_0$ exists because $\gcd(2,q)=1$ for $q>2$.) Then, the solution $(X_0,Y_0)$ is of the form $(Z_0^2,Z_0^{2s})$. By the transformation \eqref{transinv}, $(Z_0^2,Z_0^t)$ is also a solution, where $t=q^r-2s$. Then
\ie\label{qrs-ineq}
\frac{\varphi(q^r)}{2} < s \le \frac{q^r-1}{2} \implies 1\le t < q^{r-1} \le \frac{q^r-q^{r-1}}{2} = \frac{\varphi(q^r)}{2}~,
\fe
where the rightmost inequality holds for $q>2$. Clearly, $Z_0$ is a root of $\Phi_{q^r}(X)$. Since $(Z_0^2,Z_0^t)$ satisfies $\tilde{\mathsf P}(X,Y) = 0$, $Z_0$ is a root of $\tilde{\mathsf P}(X^2,X^t)$. We can write
\ie
\tilde{\mathsf P}(X^2,X^t) &= X^2(X^t-1)^2 + X^t(X^2-1)^2
\\
&=\begin{cases}
X(X-1)^2 \check{\mathsf P}_1(X)~,\qquad & t = 1~,
\\
X^2 (X-1)^2 \check{\mathsf P}_t(X)~,\qquad & t > 1~,
\end{cases}
\fe
where
\ie
\check{\mathsf P}_t(X) = \begin{cases}
X + (X+1)^2~, \qquad & t = 1~,
\\
\left(\textstyle\sum_{i=0}^{t-1}X^i\right)^2 + X^{t-2}(X+1)^2~,\qquad & t > 1~.
\end{cases}
\fe
$\check{\mathsf P}_t(X)$ is a nonzero polynomial because $\check{\mathsf P}_t(0) = 1\mod p$ for $t \ne 2$, and $2\mod p$ for $t = 2$.\footnote{Once again, this is not true for $p=2$ because $\check{\mathsf P}_t(X) = 0$ identically when $t = 2$, so there is always a solution of \eqref{poly-sys} of the form $(Z_0^2,Z_0^2)$ even for $Z_0\ne1$.\label{p=2fail2}} Since $Z_0$ is a root of $\tilde{\mathsf P}(X^2,X^t)$ and $Z_0 \ne 0,1$, it is a root of $\check{\mathsf P}_t(X)$ as well. We now use the fact that $\Phi_{q^r}(X)$ is the minimal polynomial of $Z_0$ in $\mathbb Z_p[X]$ because $p$ is a primitive root modulo $q^r$ \cite[Section~11.2.B]{cox2012}. This means $\Phi_{q^r}(X)$ must divide $\check{\mathsf P}_t(X)$. But this is impossible because
\ie
\deg_X \check{\mathsf P}_t(X) &= t + \max(t,2) - 2 + \delta_{t,1}\le 2t< \varphi(q^r)= \deg_X \Phi_{q^r}(X)~,
\fe
where in the third line, we used \eqref{qrs-ineq}. So, there is no such $Z_0$.
\end{itemize}
Therefore, when $p$ is a primitive root modulo $q^m$, then $(1,1)$ is the only solution of \eqref{poly-sys3}, and hence, $\log_p\text{GSD}$ is still given by \eqref{ZNlap-GSD}.

\end{enumerate}

To conclude, when $L_x = p^{k_x} q^m$ and $L_y = p^{k_y} q^m$, where $q$ is an odd prime such that $p$ is a primitive root modulo $q^m$, and $k_x,k_y,m\ge0$, the ground state degeneracy of the 3+1d anisotropic $\mathbb Z_p$ Laplacian model is given by
\ie
\log_p \text{GSD} = 2 \left[2 p^{\min(k_x,k_y)} - \delta_{k_x,k_y} \right]~.
\fe
When $m=0$, we see that $\log_p \text{GSD}$ scales as $4\min(L_x,L_y)$. This gives an infinite family of $L_x,L_y$ for which $\log_p \text{GSD}$ is $O(L_x,L_y)$.

Say $q$ is such that $p$ is a primitive root modulo $q^2$. Then, $p$ is a primitive root modulo $q^m$ for all $m\ge1$ (see footnote \ref{ftnt:primroot2}). Then, for $k_x = k_y = 0$ and any $m\ge0$, we see that $\log_p \text{GSD}=2$, a finite number. This gives an infinite family of $L_x,L_y$ for which $\log_p\text{GSD}$ remains finite.\footnote{In contrast, when $N=2$, $\log_2\text{GSD}$ in \eqref{3dZ2-GSD} always scales as $4L$ for any $L$. This is because, when $p=2$, the above arguments do not go through, as explained in footnotes \ref{p=2fail} and \ref{p=2fail2}.}

Note that the last conclusion relies on the existence of an odd prime $q$ such that $p$ is a primitive root modulo $q^2$. However, we do not know of a proof for general $p$. Another interesting possibility is the following. By \emph{Artin's conjecture on primitive roots} \cite{Moree2012},\footnote{The conjecture is actually stronger: the set of such $q$ has positive asymptotic density inside the set of all primes.} there are infinitely many prime $q$ such that $p$ is a primitive root modulo $q$.  (Recall from footnote \ref{ftnt:primroot2} that this does not imply that $p$ is a primitive root modulo $q^2$.) Then, choosing $L_x = L_y = q$ for all such $q$ gives another infinite family of $L_x,L_y$ for which $\log_p\text{GSD} = 2$. However, Artin's conjecture is still unproven, except under the assumption of the \emph{generalized Riemann hypothesis} \cite{Hooley1967}, which is also unproven.

\subsection{Mobility restrictions}\label{app:3dZpanisolap-mob}

We now discuss the mobility of $z$-lineons in the $xy$-plane in the 3+1d anisotropic $\mathbb Z_p$ Laplacian model. The lineons are represented as defects in the low-energy theory and their motion is implemented by operators acting at fixed time.

These operators fall into two kinds.  First, there are operators supported in a small region, e.g., the line joining the two points.  Second, there are also situations where the operator spans over $O(L_x,L_y)$ sites.  Operators of the second kind exist only for certain special values of $L_x,L_y$ depending on some number-theoretic properties of $L_x,L_y$, whereas the first kind exist for all $L_x,L_y$. In particular, only the first kind exist on an infinite square lattice. (See the discussion in \cite{Gorantla:2022eem,Gorantla:2022ssr}.)

As an example of the second kind of operator, consider $L_x = L_y = q^m$, where $q>2$ is a prime such that $p$ is a primitive root modulo $q^m$. In Appendix \ref{app:3dZpanisolap-gsd}, we showed that for $L_x = L_y = q^m$, where $q>2$ is a prime such that $p$ is a primitive root modulo $q^m$, the ground state degeneracy is given by $\log_p \text{GSD} = 2$. It follows that $|\Jac(C_{q^m} \times C_{q^m},p)| = p$, or equivalently, $\Jac(C_{q^m} \times C_{q^m},p) = \mathbb Z_p$. Therefore, in this case, the only selection imposed by the $\Jac(C_{q^m} \times C_{q^m},p)$ time-like symmetry is that the total charge of the defects is conserved modulo $p$. In particular, a $z$-lineon can move anywhere within the $xy$-plane when $L_x = L_y = q^m$. However, we show below that a $z$-lineon is completely immobile when $\Gamma$  is an infinite square lattice. This means, the operator that moves a $z$-lineon on the 2d torus graph $C_{q^m} \times C_{q^m}$ must be of the second kind.

We now show that a $z$-lineon is completely immobile on an infinite square lattice. In fact, we show that any finite configuration of $z$-lineons is completely immobile (except in some trivial cases) as long as their charges and the separations between them are fixed during the motion, i.e., we allow only ``rigid'' motion. Without this restriction, the groups of lineons can move.  We will not discuss this motion.

Our analysis will be similar to the analogous discussion in \cite{Gorantla:2022ssr}.  The main difference between them is that here our variables are in $\mathbb Z_p$ and therefore various properties of the polynomials will depend on $p$.

Consider $n$ $z$-lineons, with charges $q_i$ and positions $(x_i,y_i)$ for $i=1,\ldots,n$, described by the defect
\ie\label{poly-def}
\exp\left[ \frac{2\pi i}{p} \sum_{\tau} \sum_{i=1}^n q_i m_\tau(\tau+\tfrac12,x_i,y_i) \right]
\fe
(Since they are $z$-lineons, we can assume without loss of generality that they all have the same $z$ coordinate, and omit writing it.) They can move ``rigidly'' by $(x_0,y_0)\ne(0,0)$ if there is a defect of the form
\ie\label{poly-defmove}
&\exp\left[ \frac{2\pi i}{p} \sum_{\tau<0} \sum_{i=1}^n q_i m_\tau(\tau+\tfrac12,x_i,y_i) \right] \times \exp\left[ \frac{2\pi i}{p}\sum_{j=1}^{l} s_j m(0,x_j,y_j) \right]
\\
&\qquad \times \exp\left[ \frac{2\pi i}{p} \sum_{\tau \ge 0} \sum_{i=1}^n q_i m_\tau(\tau+\tfrac12,x_i+x_0,y_i+y_0) \right]~.
\fe
This defect is gauge invariant if and only if
\ie\label{gauge-inv}
\sum_{i=1}^n q_i \left[k(0,x_i+x_0,y_i+y_0) - k(0,x_i,y_i) \right] = \sum_{j=1}^{l} s_j (\Delta_x^2 + \Delta_y^2) k(0,x_j,y_j) \mod p~,
\fe
for any integer gauge parameter $k$ in \eqref{ZNanisolap-gaugesym}.

Using a \emph{formal Laurent power series}
\ie
\hat k(X,Y) = \sum_{(x,y)\in\mathbb Z^2} k(0,x,y) X^{-x} Y^{-y}~,
\fe
associated with the gauge parameter $k(0,x,y)$, the condition \eqref{gauge-inv} can be written as
\ie\label{poly-mobcond}
(X^{x_0} Y^{y_0} - 1) \mathsf Q(X,Y) = \mathsf S(X,Y) \mathsf P(X,Y) \mod p~,
\fe
where
\ie\label{disclap-Laupoly}
\mathsf P(X,Y) = (X-2+X^{-1}) + (Y-2+Y^{-1})~,
\fe
is the \emph{Laurent polynomial} (i.e., an element of $\mathbb Z_p[X,X^{-1},Y,Y^{-1}]$) associated with the discrete Laplacian operator $\Delta_x^2 + \Delta_y^2$, and
\ie
\mathsf Q(X,Y) = \sum_{i=1}^n q_i X^{x_i} Y^{y_i}~,\qquad \mathsf S(X,Y) = \sum_{j=1}^l s_j X^{x_j} Y^{y_j}~,
\fe
are also Laurent polynomials. The coefficients and monomials in $\mathsf Q(X,Y)$ and $\mathsf S(X,Y)$ are obtained from the defect \eqref{poly-defmove}.

If there is a Laurent polynomial $\mathsf R(X,Y)$ such that $\mathsf Q(X,Y) = \mathsf R(X,Y) \mathsf P(X,Y)$, then \eqref{poly-mobcond} can be trivially satisfied by choosing $\mathsf S(X,Y) = (X^{x_0} Y^{y_0} - 1)\mathsf R(X,Y)$. However, in this case, the defect \eqref{poly-def} can end at $\tau = 0$ as follows:
\ie
\exp\left[ \frac{2\pi i}{p} \sum_{\tau<0} \sum_{i=1}^n q_i m_\tau(\tau+\tfrac12,x_i,y_i) \right] \times \exp\left[-\frac{2\pi i}{p}\sum_{j'=1}^{l'} r_{j'} m(0,x_{j'},y_{j'}) \right]
\fe
where $r_{j'}$'s and $(x_{j'},y_{j'})$'s are obtained from $\mathsf R(X,Y) = \sum_{j'=1}^{l'} r_{j'}X^{x_{j'}} Y^{y_{j'}}$. Therefore, in this case, the defect \eqref{poly-defmove} describes the annihilation of the $n$ $z$-lineons at their original positions and their creation at positions displaced by $(x_0,y_0)$ at time $\tau = 0$.

A more interesting situation occurs for a defect like \eqref{poly-defmove} when $\mathsf Q(X,Y)$ cannot be written as $\mathsf R(X,Y)\mathsf P(X,Y)$ for any Laurent polynomial $\mathsf R(X,Y)$. Then, to satisfy \eqref{poly-mobcond}, $\mathsf P(X,Y)$ and $X^{x_0} Y^{y_0} - 1$ must share a nontrivial factor.\footnote{By \emph{nontrivial factor} we mean a nonconstant Laurent factor that is not a Laurent monomial.} Let us show that this cannot happen. In the following, it is crucial that $(x_0,y_0) \ne (0,0)$.

First, note that $\mathsf P(X,Y)$ is nonconstant and irreducible up to a monomial in $\mathbb Z_p[X,X^{-1},Y,Y^{-1}]$ for any odd prime $p$.\footnote{A polynomial in $\mathbb Z_p[X,Y]$ is said to be \emph{irreducible} if it cannot be written as a product of two nonconstant polynomials. A Laurent polynomial $\mathsf F(X,Y)$ in $\mathbb Z_p[X,X^{-1},Y,Y^{-1}]$ is said to be \emph{irreducible up to a monomial} if $X^a Y^b \mathsf F(X,Y)$ is an irreducible polynomial for some $a,b\in\mathbb Z$. For example, $\mathsf P(X,Y)$ is irreducible up to a monomial because $\tilde{\mathsf P}(X,Y) = XY\mathsf P(X,Y)$, given by \eqref{disclap-poly}, is an irreducible polynomial. The irreducibility of $\tilde{\mathsf P}(X,Y)$ for any prime $p>6$ follows from \cite[Corollary~3]{GAO2003}. It is easy to verify by hand, or in $\mathsf{Mathematica}$, that it is irreducible even for $p=3,5$. It is, however, not irreducible for $p=2$ because $\tilde{\mathsf P}(X,Y) = (XY+1)(X+Y) \mod 2$. This is one way of seeing why a dipole of $z$-lineons separated in $(1,\pm1)$ direction can move in $(1,\mp1)$ direction in the 3+1d anisotropic $\mathbb Z_2$ Laplacian model.}  So, all we need to show is that $X^{x_0} Y^{y_0} - 1$ is not a multiple of $\mathsf P(X,Y)$ in $\mathbb Z_p[X,X^{-1},Y,Y^{-1}]$.

Let $p^k$ be the largest power of $p$ that divides both $x_0$ and $y_0$, i.e., $x_0' = x_0/p^k$ and $y_0' = y_0/p^k$ are integers and $d = \gcd(x_0',y_0')$ is not divisible by $p$. Then, in $\mathbb Z_p[X,X^{-1},Y,Y^{-1}]$, we have
\ie
X^{x_0} Y^{y_0} - 1 = (X^{x_0'} Y^{y_0'} - 1)^{p^k} = \left[ (X^{x_0''} Y^{y_0''} - 1) \mathsf T(X,Y) \right]^{p^k}~,
\fe
where $x_0'' = x_0'/d$, $y_0'' = y_0'/d$, and $\mathsf T(X,Y) = \sum_{c=0}^{d-1} (X^{x_0''} Y^{y_0''})^c$. Now, $\mathsf T(1,1) = d \ne 0 \mod p$, whereas $\mathsf P(1,1) = 0$, so $\mathsf T(X,Y)^{p^k}$ is not a multiple of $\mathsf P(X,Y)$. Since $\gcd(x_0'',y_0'')=1$, the factor $X^{x_0''} Y^{y_0''} - 1$ is nonconstant and irreducible up to a monomial for any $p$ \cite{GAO2001}. So $(X^{x_0''} Y^{y_0''} - 1)^{p^k}$ is also not a multiple of $\mathsf P(X,Y)$. Therefore, $X^{x_0} Y^{y_0} - 1$ is not a multiple of $\mathsf P(X,Y)$.

To conclude, a finite set of $z$-lineons cannot move ``rigidly'' in the $xy$-plane in the 3+1d anisotropic $\mathbb Z_p$ Laplacian model, unless they can be annihilated.

\section{$\mathbb Z_N$ Laplacian model on a graph}\label{sec:ZNlap}
In this appendix, we analyze a gapped fracton model on a simple, connected, undirected spatial graph $\Gamma$. We refer to it as the $\mathbb{Z}_N$ Laplacian model because the theory is defined using the discrete Laplacian operator $\Delta_L$ on the graph $\Gamma$. The anisotropic $\mathbb{Z}_N$ Laplacian model in Section \ref{sec:ZNanisolap} is an anisotropic extension of this $\mathbb{Z}_N$ Laplacian model by adding another direction.

\subsection{Hamiltonian}\label{sec:ZNlap-Ham}
In the Hamiltonian formulation of the $\mathbb Z_N$ Laplacian model, there are a $\mathbb Z_N$ variable $U_i$ and its conjugate variable $V_i$, i.e. $U_iV_i=e^{2\pi i/N}V_i U_i$, on every site of the graph $\Gamma$ where $i$ labels the sites.
The Hamiltonian is
\ie\label{ZNlap-Ham}
H = -\gamma_1 \sum_i G_i + \text{h.c.}~,
\fe
where
\ie
G_i = \prod_{j:\langle i, j \rangle\in \Gamma} V_i V_j^\dagger ~.
\fe
Here, $\langle i,j\rangle$ means $i$ and $j$ are connected by an edge in the graph $\Gamma$.

Since all the $G_i$s commute, the ground states satisfy $G_i=1$ for all $i$ and the excitations are violations of $G_i=1$. We could take the limit $\gamma_1\rightarrow\infty$, in which case, the Hilbert space consists of only the ground states and the Hamiltonian is trivial. The Euclidean presentation of this model in this limit will be discussed later in Appendix \ref{app:Euclid_pre}.

We are particularly interested in those operators that commute with the Hamiltonian \eqref{ZNlap-Ham} and act nontrivially on its ground states. They are the global symmetry operators of the model in the low energy limit.

The electric symmetry operators are $V_i$, which trivially commute with the Hamiltonian. Since the ground states satisfy $G_i=1$, some of these operators are equivalent when acting on the ground states. The independent symmetry operators are
\ie\label{eq:tildeW}
\tilde W_\lambda=\prod_i V_i^{\lambda(i)}~,
\fe
where $\lambda(i)$ takes the form \eqref{graph-rep}
\ie
\lambda(i)=\sum_{a=1}^\mathsf{N} p_a (Q^{-1})_{ai}~,\qquad p_a\sim p_a+\text{gcd}(N,r_a)~.
\fe
Let us explain the identification on $p_a$. We have $p_a\sim p_a+N$ because $V_i^N=1$. We also have $p_a\sim p_a+r_a$ because $r_a(Q^{-1})_{ai} =\sum_j L_{ij}P_{aj}$ and $\prod_i V_i^{L_{ij}}=1$ when acting on the ground states. Combining the two identifications, we get $p_a\sim p_a+\text{gcd}(N,r_a)$.
The symmetry operators generate a $\Jac(\Gamma,N)$ electric symmetry.

The magnetic symmetry operators are
\ie\label{eq:W}
W_{\tilde \lambda}=\prod_i U_i^{\tilde \lambda(i)}~,
\fe
where $\tilde \lambda(i)$ obeys $\Delta_L\tilde \lambda(i)=0\mod N$, and the most general solution takes the form \eqref{graph-dischar}
\ie
\tilde \lambda(i) = \sum_{a=1}^\mathsf{N} \frac{NQ_{ia}\tilde p_a}{\text{gcd}(N,r_a)}~,\qquad \tilde p_a\sim \tilde p_a+\text{gcd}(N,r_a)~.
\fe
The symmetry operators generate a $\Jac(\Gamma,N)$ magnetic symmetry.

A convenient basis of electric and magnetic space-like symmetry operators is given by
\ie\label{ZNlap-ops-Hamiltonian}
&\tilde W(a) = \prod_i V_i^{ (Q^{-1})_{ai} }~,
\\
&W(a) = \prod_i U_i^{\frac{N}{\gcd(N,r_a)}  Q_{ia} }~.
\fe
for $a = 1,\ldots,\mathsf N$. Both $W(a)$ and $\tilde W(a)$ are $\mathbb Z_{\gcd(N,r_a)}$ operators. They satisfy the commutation relations
\ie\label{eq:commutation_relation_ZNlap}
W(a) \tilde W(b) =  \exp\left[ \frac{2\pi i \delta_{ab}}{\gcd(N,r_a)} \right] \tilde W(b) W(a)~,\qquad a,b = 1,\ldots,\mathsf N~.
\fe
For each $a$, there is only one $b$ that has nontrivial commutation relation. So for each $a$, there is an independent $\mathbb Z_{\gcd(N,r_a)}$ Heisenberg algebra generated by $W(a)$ and $\tilde W(a)$, leading to a ground state degeneracy
\ie
\text{GSD} = \prod_{a=1}^\mathsf{N} \gcd(N,r_a) = |\Jac(\Gamma,N)|~.
\fe

\subsection{Euclidean presentation}
\label{app:Euclid_pre}

We now discuss the Euclidean presentation of the $\mathbb{Z}_N$ Laplacian model. We place the theory on a Euclidean spacetime lattice  $C_{L_\tau}\times\Gamma$, where $\Gamma$ is the spatial slice. We use $(\tau,i)$ to label a site in the spacetime lattice. The integer $BF$-action of the $\mathbb Z_N$ Laplacian theory is
\ie\label{ZNlap-modVill-action}
S = \frac{2\pi i}{N} \sum_{\tau,i} \tilde m (\tau+\tfrac12,i) \left[ \Delta_\tau m(\tau,i) - \Delta_L m_\tau(\tau+\tfrac12,i) \right]~,
\fe
where the integer fields $\tilde m$ and $(m_\tau, m)$ have an integer gauge symmetry
\ie\label{eq:gauge_sym_app}
&\tilde m \sim \tilde m + N \tilde k~,
\\
&m_\tau \sim m_\tau + \Delta_\tau k + N q_\tau~,
\\
&m \sim m + \Delta_L k + N q~,
\fe
where $k$, $\tilde k$, and $(q_\tau,q)$ are integers. (Note that, when working modulo $N$, the last line of \eqref{eq:gauge_sym_app} is precisely the equivalence relation discussed in \eqref{graph-equiv}.) The integer $BF$-action \eqref{ZNlap-modVill-action} describes the ground states of the Hamiltonian \eqref{ZNlap-Ham}.

\subsubsection{Ground state degeneracy}
We can count the number of ground states by counting the number of solutions to the ``equations of motion'' of $(m_\tau,m)$:
\ie\label{ZNlap-eom}
\Delta_\tau \tilde m = 0 \mod N~,\qquad \Delta_L \tilde m = 0 \mod N~.
\fe
The first equation implies that $\tilde m(\tau,i)$ is independent of $\tau$. Then, as discussed in Section \ref{sec:graph-disclap}, the general solution is
\ie\label{ZNlap-gen-sol}
\tilde m(i) = \sum_{a=1}^\mathsf{N} \frac{NQ_{ia} p_a}{\gcd(N,r_a)}~,\qquad p_a\sim p_a+\text{gcd}(N,r_a)~.
\fe
Therefore, the ground state degeneracy is
\ie\label{ZNlap-gsd}
\text{GSD} = \prod_{a=1}^\mathsf{N} \gcd(N,r_a) = |\Jac(\Gamma,N)|~.
\fe

\subsubsection{Global symmetry}
The above ground state degeneracy can also be obtained from the (space-like) global symmetry. There are electric (space-like and time-like) and magnetic (space-like) global symmetries, whose groups are both
\ie
\Jac(\Gamma,N) = \prod_{a=1}^\mathsf{N} \mathbb Z_{\gcd(N,r_a)}~.
\fe

The electric global symmetry acts as
\ie
(m_\tau,m) \rightarrow (m_\tau,m) + (\lambda_\tau,\lambda)~,
\fe
where $(\lambda_\tau,\lambda)$ is a flat $\mathbb Z_N$ gauge field, i.e., $\Delta_\tau \lambda - \Delta_L \lambda_\tau = 0 \mod N$. Using $k$, we can set $\lambda_\tau(\tau+\tfrac12,i)|_{\tau\ne0} = 0 \mod N$. Then, by flatness, we have $\Delta_\tau \lambda = 0 \mod N$. This in turn implies that $\Delta_L \lambda_\tau(\tau+\tfrac12,i)|_{\tau=0} = 0 \mod N$, which is the discrete Laplace equation \eqref{graph-disclap}.

The remaining time-independent gauge freedom, $\lambda(i) \sim \lambda(i) + \Delta_L k(i)$, is precisely the equivalence relation in \eqref{graph-equiv}. Therefore, we can gauge fix $\lambda(i)$ to
\ie\label{ZNlap-elecsym}
\lambda(i) = \sum_{a=1}^\mathsf{N} p_a (Q^{-1})_{ai}~,
\fe
where $p_a = 0,\ldots,\gcd(N,r_a)-1$. Since $\lambda_\tau(\tau+\tfrac12,i)|_{\tau=0}$ satisfies the discrete Laplace equation, the most general solution is \eqref{graph-dischar}
\ie\label{ZNlap-elecsym-time}
\lambda_\tau(\tau+\tfrac12,i)|_{\tau=0} = \sum_{a=1}^\mathsf{N} \frac{N Q_{ia} p_{\tau,a}}{\gcd(N,r_a)}~,
\fe
where $p_{\tau, a} = 0,\ldots,\gcd(N,r_a)-1$. The parameters $p_{\tau, a}$ and $p_a$ generate the electric time-like and space-like global symmetries respectively.

The magnetic space-like global symmetry acts as
\ie\label{ZNlap-magsym}
\tilde m(\tau+\tfrac12,i) \rightarrow \tilde m(\tau+\tfrac12,i) + \tilde \lambda(i)~,\qquad \tilde \lambda(i) = \sum_{a=1}^\mathsf{N} \frac{N Q_{ia} \tilde p_a}{\gcd(N,r_a)}~,
\fe
and $\tilde p_a = 0,\ldots,\gcd(N,r_a)-1$.

A convenient basis of electric and magnetic space-like symmetry operators is given by
\ie\label{ZNlap-ops}
&\tilde W(a) = \exp\left[ \frac{2\pi i}{N} \sum_i (Q^{-1})_{ai} \tilde m(\tau+\tfrac12,i) \right]~,
\\
&W(a) = \exp\left[ \frac{2\pi i}{\gcd(N,r_a)} \sum_i m(\tau,i) Q_{ia} \right]~.
\fe
for $a = 1,\ldots,\mathsf N$. These operators are the low energy counterpart of the operators in \eqref{ZNlap-ops-Hamiltonian}. The commutation relation \eqref{eq:commutation_relation_ZNlap} can now be understood as a mixed 't Hooft anomaly between the electric and magentic space-like symmetries.

\subsubsection{Time-like symmetry and fractons}

The $\mathbb{Z}_N$ Laplacian model has defects that extend in the time direction, such as
\ie\label{ZNlap-def}
W_\tau(i) = \exp\left[ \frac{2\pi i}{N} \sum_{\tau} m_\tau(\tau+\tfrac12,i) \right]~.
\fe
This describes the world-line of an infinitely heavy particle of unit charge at position $i \in \Gamma$.

Below we discuss the time-like global symmetry that acts on these defects.
The electric time-like symmetry acts as
\ie
m_\tau(\tau+\tfrac12,i) \rightarrow m_\tau(\tau+\tfrac12,i) + \delta_{\tau,0} \sum_{a=1}^\mathsf{N} \frac{N Q_{ia} p_{\tau, a}}{\gcd(N,r_a)}~.
\fe
Therefore, two defects at sites $i$ and $i'$ carry the same time-like charges, or equivalently, a particle can hop from $i$ to $i'$, if and only if
\ie\label{ZNlap-mob}
Q_{ia} = Q_{i'a} \mod \gcd(N,r_a)~,\qquad a=1,\ldots,\mathsf N~.
\fe
Indeed, when this condition holds, the defect that ``moves'' a particle from $i$ to $i'$ at time $\tau=0$ is given by
\ie
&\exp\left[ \frac{2\pi i}{N} \sum_{\tau<0} m_\tau(\tau+\tfrac12,i) \right]\exp\left[ -\frac{2\pi i}{N} \sum_{a,j} \left( \frac{Q_{ia} - Q_{i'a}}{\gcd(N,r_a)} \right) \tilde r_a P_{aj} m(0,j) \right]
\\
&\qquad \qquad \times\exp\left[ \frac{2\pi i}{N} \sum_{\tau\ge0} m_\tau(\tau+\tfrac12,i') \right]~,
\fe
where $\tilde r_a$ is an integer such that $\tilde r_a r_a = \gcd(N,r_a)\mod N$.

While the selection rule \eqref{ZNlap-mob} is not very intuitive, it leads to strong mobility constraints in the special case where the spatial lattice is a square lattice (i.e., $\Gamma$ is a 2d torus graph $C_{L_x} \times C_{L_y}$).  This will be shown in Section \ref{app:2dZNlap}. In particular, under some mild conditions, the particles are completely immobile, i.e., they are fractons.

\subsubsection{Robustness}\label{app:ZNlaprobust}

We now discuss the robustness of the low-energy limit of the $\mathbb Z_N$ Laplacian model. The only operators that act nontrivially on the ground states are $W(a)$ and $\tilde W(a)$ of \eqref{ZNlap-ops}.
$W(a)$ is an extended operator with support spanning over the entire graph.
In contrast, $\tilde W(a)$ can be written as a product of local operators of the form $e^{\frac{2\pi i}{N} \tilde m(\tau+\frac12,i)}$.
(In \eqref{ZNlap-ops} we defined $\tilde W(a)$ in such a way that its commutation relation \eqref{eq:commutation_relation_ZNlap} with the extended operator $W(a)$ is simple.)
Since these local operators act nontrivially in the space of ground states, the low-energy limit of the model is not robust.

\subsection{Examples}

\subsubsection{$\Gamma = C_{L_x}$}
Let $\Gamma$ be a cycle graph, i.e., $\Gamma = C_{L_x}$, where $L_x$ is the number of sites in the cycle. The operator $\Delta_L$ associated with the Laplacian matrix of $\Gamma$ is the same as the standard Laplacian operator $\Delta_x^2$ in the $x$-direction.

In this case, the $\mathbb{Z}_N$ Laplacian model simplifies to the 1+1d rank-2 $\mathbb Z_N$ tensor gauge theory of \cite{Gorantla:2022eem}. Indeed, the diagonal entries in the Smith normal form of $L$ are
\ie
r_a = \begin{cases}
	1~,& 1\le a < L_x -1~,\\
	L_x~,& a = L_x-1~,\\
	0~,& a = L_x~.
\end{cases}
\fe
To be more concrete, we can write $R = P L Q$, where
\ie
R= \begin{pmatrix} I_{L_x-2} & \b 0 & \b 0\\ \b 0^T & L_x & 0\\ \b 0^T & 0 & 0 \end{pmatrix}~,\qquad P = \begin{pmatrix} \tilde P &\b 0\\ \b 1^T&1 \end{pmatrix}~,\qquad Q = \begin{pmatrix} \tilde Q &\b 1\\ \b 0^T&1 \end{pmatrix}~,
\fe
where $\tilde P$ and $\tilde Q$ are $(L_x-1)\times(L_x-1)$ integer matrices given by
\ie
\tilde P_{a,x+1} = \min\{a,x+1\}~,\qquad \tilde Q_{x+1,a} = \delta_{x+1,a} - (x+1) (1-\delta_{x,L_x-2}) \delta_{a,L_x-1}~,
\fe
where $a=1,\ldots,L_x$, and $x=0,\ldots,L_x-1$. For example, for $L_x = 5$, the $4\times 4$ matrices $\tilde P$ and $\tilde Q$ are
\ie
\tilde P = \begin{pmatrix} 1 & 1 & 1 & 1\\ 1 & 2 & 2 & 2\\ 1 & 2 & 3 & 3\\1 & 2 & 3 & 4 \end{pmatrix}~,\qquad \tilde Q = \begin{pmatrix} 1 & 0 & 0 & -1\\ 0 & 1 & 0 & -2\\ 0 & 0 & 1 & -3\\ 0 & 0 & 0 & 1 \end{pmatrix}~.
\fe

The electric (time-like and space-like) and magnetic (space-like) symmetries of the $\mathbb Z_N$ Laplacian gauge theory are $\mathbb Z_N \times \mathbb Z_{\gcd(N,L_x)}$. These are in agreement with the 1+1d rank-2 $\mathbb Z_N$ tensor gauge theory of \cite{Gorantla:2022eem}.

\subsubsection{$\Gamma=C_{L_x}\times C_{L_y}$}\label{app:2dZNlap}

Let $\Gamma$ be a torus graph, i.e., $\Gamma = C_{L_x}\times C_{L_y}$, where $L_x$ and $L_y$ are the number of sites in the $x$-cycle and $y$-cycle. The operator $\Delta_L$ associated with the Laplacian matrix of $\Gamma$ is the same as the standard Laplacian operator $\Delta_x^2+\Delta_y^2$ on a square lattice.

In this case, the $\mathbb{Z}_N$ Laplacian model can be viewed as the $\mathbb{Z}_N$ version of the Laplacian $\phi$-theory or the $\mathbb{Z}_N$ version of the $U(1)$ Laplacian gauge theory discussed in \cite{Gorantla:2022mrp,Gorantla:2022ssr}. Its ground state degeneracy is the square root of that of the anisotropic $\mathbb{Z}_N$ Laplacian model, which is computed in Appendix \ref{app:3dZpanisolap-gsd}.
When $N$ is a prime, the GSD of the $\mathbb{Z}_N$ Laplacian model is thus
\ie
\log_N\text{GSD}=\text{dim}_{\mathbb{Z}_N}\frac{\mathbb{Z}_N[X,Y]}{(Y(X-1)^2+X(Y-1)^2),X^{L_x}-1,Y^{L_y}-1)}~.
\fe
The GSD depends on $L_x,L_y$ in an erratic way. There exists a sequence of $L_x,L_y$ where the $\log_N\text{GSD}\sim O(L_x,L_y)$, but there also exists a sequence where the $\text{GSD}$ stays at order 1 if $N>2$.

Like the GSD, the mobility of a particles depends on number-theoretic properties of $L_x,L_y$. Since the mobility of these particles is the same as the mobility of the $z$-lineons of the 3+1d anisotropic $\mathbb Z_N$ Laplacian model in the $xy$-plane, the analysis of Appendix \ref{app:3dZpanisolap-mob} applies here. In particular, when $N$ is an odd prime, there are infinitely many values of $L_x,L_y$ for which a single particle is completely mobile. In contrast, on an infinite square lattice, any finite set of particles is completely immobile (unless they can be annihilated), assuming they move ``rigidly.''

 When $N=2$, the $\mathbb Z_N$ Laplacian model is equivalent to two copies of a known model, the $\mathbb Z_2$ Ising plaquette model \cite{paper1}, when both $L_x$ and $L_y$ are even, and only one copy when $L_x$ or $L_y$ is odd. Therefore, the GSD and mobility restrictions of the $\mathbb Z_2$ Laplacian model are relatively simple in this case, and follow from the analysis in Appendix \ref{app:3dZ2anisolap}.

Let us contrast the $\mathbb{Z}_N$ Laplacian model with the 2+1d rank-2 $\mathbb{Z}_N$ tensor gauge theory discussed in \cite{Bulmash:2018lid,Ma:2018nhd,Oh2021,Oh2022,Pace2022}, which is another 2+1d generalization of the 1+1d rank-2 $\mathbb{Z}_N$ tensor gauge theory. These two models differ in several aspects:
\begin{enumerate}
\item The $\mathbb{Z}_N$ tensor gauge theory has a ground state degeneracy of
\ie
N^3\text{gcd}(N,L_x)\text{gcd}(N,L_y)\text{gcd}(N,L_x,L_y)~.
\fe
In particular, the GSD of $\mathbb{Z}_N$ tensor gauge theory is always bounded by $N^6$, whereas there are infinitely many $L_x,L_y$ for which $\log_N \text{GSD}$ of the $\mathbb Z_N$ Laplacian model scales as $O(L_x,L_y)$, at least when $N$ is prime.

\item Relatedly, the low-energy limit of the $\mathbb{Z}_N$ tensor gauge theory is robust, whereas the low-energy limit of the $\mathbb Z_N$ Laplacian model is not.

\item A particle in the $\mathbb{Z}_N$ tensor gauge theory can always hop by $N$ sites on an infinite square lattice,\footnote{More precisely, this is true for an electrically charged particle. Magnetically charged particles come in two flavors: $x$-lineons which can move anywhere in the $x$ direction but can only hop by $N$ sites in the $y$-direction, and similarly $y$-lineons.} whereas a particle in the $\mathbb Z_N$ Laplacian model is completely immobile on an infinite square lattice, at least when $N$ is prime.
\end{enumerate}

\bibliographystyle{JHEP}
\bibliography{laplacian,fracton}

\end{document}